\documentclass[longauth]{aa} % for the long lists of affiliations 

\usepackage[varg]{txfonts}
\usepackage{graphicx}
\usepackage{threeparttable}
\usepackage{multirow}

\usepackage{hyperref}
\usepackage[switch]{lineno}
\usepackage{ulem}
\usepackage{color}

\titlerunning{MAGIC observations in the vicinity of SNR G106.3+2.7}
\begin{document} 

\title{
MAGIC observations provide compelling evidence of the hadronic multi-TeV emission from the putative PeVatron SNR G106.3+2.7
}
%\subtitle{}
\authorrunning{MAGIC Collaboration et al.}

\author{
%\normalsize
\fontsize{9.5pt}{3pt}\selectfont
MAGIC Collaboration:
H.~Abe\inst{1} \and
S.~Abe\inst{1} \and
V.~A.~Acciari\inst{2} \and
I.~Agudo\inst{3} \and
T.~Aniello\inst{4} \and
S.~Ansoldi\inst{5,45} \and
L.~A.~Antonelli\inst{4} \and
A.~Arbet Engels\inst{6} \and
C.~Arcaro\inst{7} \and
M.~Artero\inst{8}
% ...
}

\institute { 
%\scriptsize
\fontsize{7.5pt}{0pt}\selectfont
Japanese MAGIC Group: Institute for Cosmic Ray Research (ICRR), The University of Tokyo, Kashiwa, 277-8582 Chiba, Japan
\and Instituto de Astrof\'isica de Canarias and Dpto. de  Astrof\'isica, Universidad de La Laguna, E-38200, La Laguna, Tenerife, Spain
\and Instituto de Astrof\'isica de Andaluc\'ia-CSIC, Glorieta de la Astronom\'ia s/n, 18008, Granada, Spain
\and National Institute for Astrophysics (INAF), I-00136 Rome, Italy
\and Universit\`a di Udine and INFN Trieste, I-33100 Udine, Italy
\and Max-Planck-Institut f\"ur Physik, D-80805 M\"unchen, Germany
\and Universit\`a di Padova and INFN, I-35131 Padova, Italy
\and Institut de F\'isica d'Altes Energies (IFAE), The Barcelona Institute of Science and Technology (BIST), E-08193 Bellaterra (Barcelona), Spain
\and Technische Universit\"at Dortmund, D-44221 Dortmund, Germany
\and Croatian MAGIC Group: University of Zagreb, Faculty of Electrical Engineering and Computing (FER), 10000 Zagreb, Croatia
\and IPARCOS Institute and EMFTEL Department, Universidad Complutense de Madrid, E-28040 Madrid, Spain
\and Centro Brasileiro de Pesquisas F\'isicas (CBPF), 22290-180 URCA, Rio de Janeiro (RJ), Brazil
\and University of Lodz, Faculty of Physics and Applied Informatics, Department of Astrophysics, 90-236 Lodz, Poland
\and Deutsches Elektronen-Synchrotron (DESY), D-15738 Zeuthen, Germany
\and ETH Z\"urich, CH-8093 Z\"urich, Switzerland
\and Universit\`a di Pisa and INFN Pisa, I-56126 Pisa, Italy
\and Universitat de Barcelona, ICCUB, IEEC-UB, E-08028 Barcelona, Spain
\and Armenian MAGIC Group: A. Alikhanyan National Science Laboratory, 0036 Yerevan, Armenia
\and Centro de Investigaciones Energ\'eticas, Medioambientales y Tecnol\'ogicas, E-28040 Madrid, Spain
\and Department for Physics and Technology, University of Bergen, Norway
\and INFN MAGIC Group: INFN Sezione di Catania and Dipartimento di Fisica e Astronomia, University of Catania, I-95123 Catania, Italy
\and INFN MAGIC Group: INFN Sezione di Torino and Universit\`a degli Studi di Torino, I-10125 Torino, Italy
\and INFN MAGIC Group: INFN Sezione di Bari and Dipartimento Interateneo di Fisica dell'Universit\`a e del Politecnico di Bari, I-70125 Bari, Italy
\and Croatian MAGIC Group: University of Rijeka, Faculty of Physics, 51000 Rijeka, Croatia
\and Universit\"at W\"urzburg, D-97074 W\"urzburg, Germany
\and University of Geneva, Chemin d'Ecogia 16, CH-1290 Versoix, Switzerland
\and Finnish MAGIC Group: Finnish Centre for Astronomy with ESO, University of Turku, FI-20014 Turku, Finland
\and Departament de F\'isica, and CERES-IEEC, Universitat Aut\`onoma de Barcelona, E-08193 Bellaterra, Spain
\and Japanese MAGIC Group: Physics Program, Graduate School of Advanced Science and Engineering, Hiroshima University, 739-8526 Hiroshima, Japan
\and Armenian MAGIC Group: ICRANet-Armenia, 0019 Yerevan, Armenia
\and Croatian MAGIC Group: University of Split, Faculty of Electrical Engineering, Mechanical Engineering and Naval Architecture (FESB), 21000 Split, Croatia
\and Croatian MAGIC Group: Josip Juraj Strossmayer University of Osijek, Department of Physics, 31000 Osijek, Croatia
\and Japanese MAGIC Group: Department of Physics, Tokai University, Hiratsuka, 259-1292 Kanagawa, Japan
\and Universit\`a di Siena and INFN Pisa, I-53100 Siena, Italy
\and Saha Institute of Nuclear Physics, A CI of Homi Bhabha National Institute, Kolkata 700064, West Bengal, India
\and Inst. for Nucl. Research and Nucl. Energy, Bulgarian Academy of Sciences, BG-1784 Sofia, Bulgaria
\and Japanese MAGIC Group: Department of Physics, Yamagata University, Yamagata 990-8560, Japan
\and Finnish MAGIC Group: Space Physics and Astronomy Research Unit, University of Oulu, FI-90014 Oulu, Finland
\and Japanese MAGIC Group: Department of Physics, Kyoto University, 606-8502 Kyoto, Japan
\and Japanese MAGIC Group: Institute for Space-Earth Environmental Research and Kobayashi-Maskawa Institute for the Origin of Particles and the Universe, Nagoya University, 464-6801 Nagoya, Japan
\and Croatian MAGIC Group: Ru\dj{}er Bo\v{s}kovi\'c Institute, 10000 Zagreb, Croatia
\and INFN MAGIC Group: INFN Sezione di Perugia, I-06123 Perugia, Italy
\and INFN MAGIC Group: INFN Roma Tor Vergata, I-00133 Roma, Italy
\and Japanese MAGIC Group: Department of Physics, Konan University, Kobe, Hyogo 658-8501, Japan
\and also at International Center for Relativistic Astrophysics (ICRA), Rome, Italy
% ...
}
\vspace{-1.5cm}
%\author{T. Oka
%  \thanks{\email{contact.magic@mpp.mpg.de}}  
%  \and T. Saito
%  \and H. Kubo
%  \and M. Strzys
%  \and ... for the MAGIC Collaboration
%}
%\offprints{
%contact.magic@mpp.mpg.de, \\
%$^*$Corresponding authors: T.~Oka, T.~Saito, M.~Strzys\\
%}
%\date{\today}
\date{Received date / Accepted date }

% \abstract{}{}{}{}{} 
% 5 {} token are mandatory

\abstract
   {Certain types of Supernova remnants (SNRs) in our Galaxy are assumed to be PeVatrons, capable of accelerating cosmic rays (CRs) to $\sim$ PeV energies. However, conclusive observational evidence for this has not yet been found. The SNR G106.3+2.7, detected at 1--100~TeV energies by different $\gamma$-ray facilities, is one of the most promising PeVatron candidates. This SNR has a cometary shape which can be divided into a {\it head} and a {\it tail} region with different physical conditions. However, it is not identified in which region the 100~TeV emission is produced due to the limited position accuracy and/or angular resolution of existing observational data. Additionally, it remains unclear whether the origin of the $\gamma$-ray emission is leptonic or hadronic.
 }
   % Aims.
   {
   With the better angular resolution provided by these new MAGIC data compared to earlier $\gamma$-ray datasets, we aim to reveal the acceleration site of PeV particles and the emission mechanism by resolving the SNR G106.3+2.7 with 0.1$^\circ$ resolution at TeV energies.
   }
  {We observed the SNR G106.3+2.7 using the MAGIC telescopes for 121.7~hours in total after quality cuts, between May 2017 and August 2019. The analysis energy threshold is $\sim$ 0.2~TeV, and the angular resolution is 0.07--0.1$^\circ$. The $\gamma$-ray spectra of different parts of the emission are examined, benefiting from the unprecedented statistics and angular resolution at these energies provided by our new data. The measurements at other wavelengths such as radio, X-rays, GeV $\gamma$-rays and 10~TeV $\gamma$-rays are also used to model the emission mechanism precisely.
  }
   {We detected extended $\gamma$-ray emission spatially coincident with the radio continuum emission at the {\it head} and {\it tail} of SNR G106.3+2.7. The fact that we detected a significant $\gamma$-ray emission with energies above 6.0~TeV from the {\it tail} region only suggests that the emissions above 10 TeV, detected with air shower experiments (Milagro, HAWC, Tibet AS$\gamma$ and LHAASO), are emitted only from the SNR {\it tail}. Under this assumption, the multi-wavelength spectrum of the {\it head} region can be explained with either hadronic or leptonic models, while the leptonic model for the {\it tail} region is in contradiction with the emission above 10~TeV and X-rays. In contrast, the hadronic model could reproduce the observed spectrum at the {\it tail} by assuming a proton spectrum with a cutoff energy of $\sim 1$~PeV for the {\it tail} region. Such a high energy emission in this middle-aged SNR (4--10~kyr) can be explained by considering the scenario that protons escaping from the SNR in the past interact with surrounding dense gases at present.
   }
   {The $\gamma$-ray emission region detected with the MAGIC telescopes in the SNR G106.3+2.7 is extended and spatially coincident with the radio continuum morphology. The multi-wavelength spectrum of the emission from the {\it tail} region suggests proton acceleration up to $\sim$ PeV, while the emission mechanism of the {\it head} region can be both hadronic or leptonic.
}


\maketitle
%
%-------------------------------------------------------------------

\section{Introduction} \label{sec:intro}
% general
It is widely assumed that cosmic rays (CRs) are accelerated to energies up to $\sim$ PeV at a shock wave in supernova remnants (SNRs) in our Galaxy \citep[see, e.g.,][and references therein]{Blasi:2013rva}.
The detection of a non-thermal synchrotron X-ray emission in a variety of SNRs~\citep[e.g.,][]{Koyama:1995rr} suggests
an acceleration of electrons up to hundreds of TeV energies, while the GeV $\gamma$-ray emission from SNRs IC 443, W44 and W51C observed with AGILE/\textit{Fermi}-LAT provides evidence for proton acceleration in SNRs~\citep{Ackermann:2013wqa, Jogler:2016lav, AGILE:2011tzq, Cardillo:2016rev}.
However, so far there is no conclusive observation of a SNR accelerating hadronic particles up to $\sim$ PeV energies, so called PeVatron.

% Boomerang (General)
The SNR G106.3+2.7 was first discovered by the northern galactic plane survey at 408 MHz with the Dominion Radio Astrophysical Observatory~\citep[DRAO;][]{Joncas_1990}.
The SNR has a comet-shaped radio morphology, with a bright circular {\it head} region and a dimmer {\it tail} region elongated to the southeast.
The double-component structure of SNR G106.3+2.7 was also observed at a frequency of 2.7 GHz~\citep{1990A&AS...85..691F}.
The {\it tail} region has a marginally softer spectrum, with $\alpha = 0.70 \pm 0.07$, than the {\it head} region, $\alpha = 0.49 \pm 0.05$~\citep{Pineault_2000}, with $\alpha$ being the index of flux density $S_{\nu} \propto \nu^{-\alpha}$.
Although the origin of the comet-shaped morphology is not well understood, HI observations suggest this is due to the distribution of the surrounding gases~\citep{Kothes_2001}.
Association of HI and molecular materials with SNR G106.3+2.7 suggests that the distance is 800~pc~\citep{Kothes_2001}, while the estimation from X-ray absorption indicates that it is  3~kpc~(\citet{Halpern:2000qh}).
At the north of the {\it head} region, there is an off-centered pulsar wind nebula (PWN) dubbed ``Boomerang''.
It is powered by the pulsar PSR J2229+6114, which has a characteristic age of 10\,kyr and a spin-down luminosity of $2.2 \times 10^{37}\,\mathrm{erg s^{-1}}$~\citep{Halpern:2001fc}.
The spectrum of the PWN shows a spectral break at 4.3 GHz attributed to synchrotron cooling~\citep{2006ApJ...638..225K}.

In the X-ray band, this SNR was recently studied using the archival Chandra, XMM-Newton and Suzuku data. 
Besides the bright emission from the PWN, non-thermal X-ray emission has been found in both the {\it head} and the {\it tail} regions~\citep{Ge:2020uft, Fujita:2021fzh}.
\cite{Fujita:2021fzh} claims that the emission in both regions is generated by electrons originating in the PWN, while \cite{Ge:2020uft} argue that the {\it tail} emission is more likely due to the electrons accelerated in the shock of the SNR.

% Boomerang (gamma ray)
\textit{Fermi}-LAT detected pulsed GeV emission from PSR J2229+6114~\citep{Abdo:2009nm}, which is associated with the previously unidentified EGRET source 3EG J2227+6122~\citep{Hartman:1999fc}.
After subtracting the emission from the pulsar, \cite{Xin:2019xeb} found a steady GeV emission in the range of 3--500~GeV from the Fermi-LAT data at the {\it tail} region.
The emission region was better described with a 0.25$^\circ$ radius disk than a point-like source.
In addition, \cite{Fang:2022uge} carefully reanalyzed the \textit{Fermi}-LAT data after removing the effect of the pulsed emissions from Boomerang and then obtained consistent results with those of~\cite{Xin:2019xeb}.
In \cite{Acciari:2009zz}, VERITAS reported a detection of extended very-high-energy (VHE) $\gamma$-ray emission in the range of 630\,GeV--17\,TeV from the {\it tail} region. It is $\sim$ 0.4$^\circ$ away from the position of PSR J2229+6114, and dubbed VER J2227+608. The shape of the emission region can be characterised with an elongated two-dimensional Gaussian with $0.27\pm0.05$ ($0.18\pm0.03$)$^\circ$ extent in the major (minor) axis. 
The VHE spectrum measured with VERITAS is well fitted by a single power-law $dN/dE$ $=$ $N_{0}$ ($E$/3\,TeV)$^{-\Gamma}$ with an index of $\Gamma$ = 2.29 $\pm$ 0.33$_{\mathrm{stat}}$ $\pm$ 0.30$_{\mathrm{sys}}$ and a flux of $N_{0}$ = (1.15 $\pm$ $0.27_{\mathrm{stat}}$ $\pm$ $0.35_{\mathrm{sys}}$) $\times$ $10^{-13}$ cm$^{-2}$ s$^{-1}$ TeV$^{-1}$~\citep{Acciari:2009zz}.
Moreover, the GeV emission reported by \cite{Xin:2019xeb} is in fact consistent within uncertainties with VER J2227+608 in position, size and spectrum.
The extended $\gamma$-ray emission spatially coincides with molecular clouds traced by $^{12}$CO ($J=1-0$) emission~\citep{Heyer_1998, Kothes_2001}, favoring a hadronic origin of the $\gamma$-ray emission.

The Milagro collaboration reported on the detection of extended VHE $\gamma$-ray emission above 20~TeV from the vicinity of the SNR. It is labelled C4~\citep{Abdo:2007ad} or MGRO J2228+61~\citep{Abdo:2009ku, 2009ATel.2172....1G}.
HAWC, Tibet AS$\gamma$, and LHAASO collaborations also reported on the detection of VHE $\gamma$-ray emission above tens of TeV from the same region~\citep{Albert:2020ngw, Tibet2021NatAs, Cao2021}.
HAWC and Tibet AS$\gamma$ results suggest a power law spectrum without a cutoff and the spectral indices are $2.25 \pm 0.23_{\rm stat}$ and $3.17 \pm 0.63_{\rm stat}$, respectively. Due to the limited angular resolution of air-shower type detectors, it is not clear if this emission comes from the {\it head} region or {\it tail} region, while it is significantly offset from the position of PSR J2229+6114. 
This very high energy emission above tens of TeV provides a lower limit on the maximum energy of the particles accelerated in this object. 
If the emission process is leptonic, an exponential cutoff energy of the electron must be higher than $270$~TeV~\citep{Albert:2020ngw} or 190~TeV~\citep{Tibet2021NatAs}, while if it is hadronic, the maximum proton energy should be higher than $800$~TeV~\citep{Albert:2020ngw} or 500~TeV ~\citep{Tibet2021NatAs}.
While it is certain that particles are accelerated to hundreds of TeV in this complex region, it is still inconclusive whether the emission originates from hadronic, leptonic or a combined process, as well as whether parent particles are accelerated in the SNR blast wave or the PWN complex.
It should also be noted that the SNR with an age of 4--10~kyr is not expected to accelerate particles to such high energies.
% About this work
In this paper, we study this complex region using deep observations with the MAGIC telescopes, which provide a better angular resolution than the ones of previous $\gamma$-ray observations of G106.3+2.7.
In Sect.~\ref{section:MAGICAna}, we describe the observations that we performed with the MAGIC telescopes.
In Sect.~\ref{section:MAGICResults}, we show the observed morphology and spectral properties.
In Sect.~\ref{section:Modelling}, we show the spectral modelling results for the multi-wavelength spectrum.
The origin of the $\gamma$-ray emission is discussed in Sect.~\ref{section:Discussion}.
We summarize the results and discuss on the future perspectives in Sect.~\ref{section:Summary}.

%--------------------------------------------------------------------
\section{Observation and data reduction} 
\label{section:MAGICAna}

% About MAGIC telescope
The MAGIC (Major Atmospheric Gamma Imaging Cherenkov) telescopes consist of two 17\,m diameter imaging Cherenkov telescopes located at 2200\,m altitude above sea level at the Observatorio del Roque de los Muchachos on the Canary island La Palma, Spain ($28.76^\circ$ N; $17.89^\circ$ W).
The MAGIC stereoscopic system is able to detect $(0.76 \pm 0.04)\%$ of the Crab Nebula flux above 210 GeV at 5$\sigma$ significance in 50 hours of observations at medium (30$^\circ$--45$^\circ$) zenith angles~\citep{Aleksic:2014lkm}.

% About Boomerang observation
VER J2227+608 was observed from May 2017 to August 2019, for 183.7 hours, at zenith angles between 30$^\circ$ and 50$^\circ$, yielding an analysis energy threshold of $\sim$ 0.2~TeV. The MAGIC angular resolution, characterised by the point spread function (PSF), for this analysis was estimated to be 0.084$^\circ$ (68$\%$ containment radius) at $E$ $>$ 0.2~TeV and 0.072$^\circ$ at $E$ $>$ 1\,TeV, which is the best angular resolution among the previous $\gamma$-ray observations for this object (e.g. 68$\%$ containment radius of the observation with the VERITAS telescope performed in 2009 is 0.11$^\circ$).

% About camera pointing
To estimate the background simultaneously, all observations were performed in wobble mode~\citep{Fomin:1994aj} at three positions (RA $=$ 336.31$^\circ$, DEC $=$ 61.40$^\circ$; RA = 338.25$^\circ$, DEC $=$ 61.06$^\circ$; RA = 336.66$^\circ$, DEC $=$ 60.42$^\circ$) with an offset of 0.57$^\circ$ from the position (RA $=$ 337.05$^\circ$, DEC $=$ 60.96$^\circ$), which is close to the center of VER J2227+608 (RA $=$ 337.0$^\circ$, DEC $=$ 60.8$^\circ$).

% About quality cut
The data analysis was performed with the MAGIC standard analysis package~\citep{Zanin:2013oib}. 
The data selection was based mainly on the transmission of the atmosphere monitored with a LIDAR system~\citep{Fruck:2014mja}.
In this analysis we only selected data with an atmospheric transmission above $85\%$ of the optimum.
After quality cuts, the total dead time corrected observation time is 121.7\,hours.
We used the wobble map method~\citep[e.g.,][]{Vovk:2018grt} for estimating backgrounds.
To cross-check the results obtained with the MAGIC standard analysis package, we used the SkyPrism package~\citep{Vovk:2018grt}, which includes independent methods to compute the instrument response functions and estimate the energy spectra using a spatial, maximum likelihood fit. Both results are in good agreement.

%--------------------------------------------------------------------
\section{Results} \label{section:MAGICResults}

\begin{figure*}
  \centering
  \includegraphics[width=16cm]{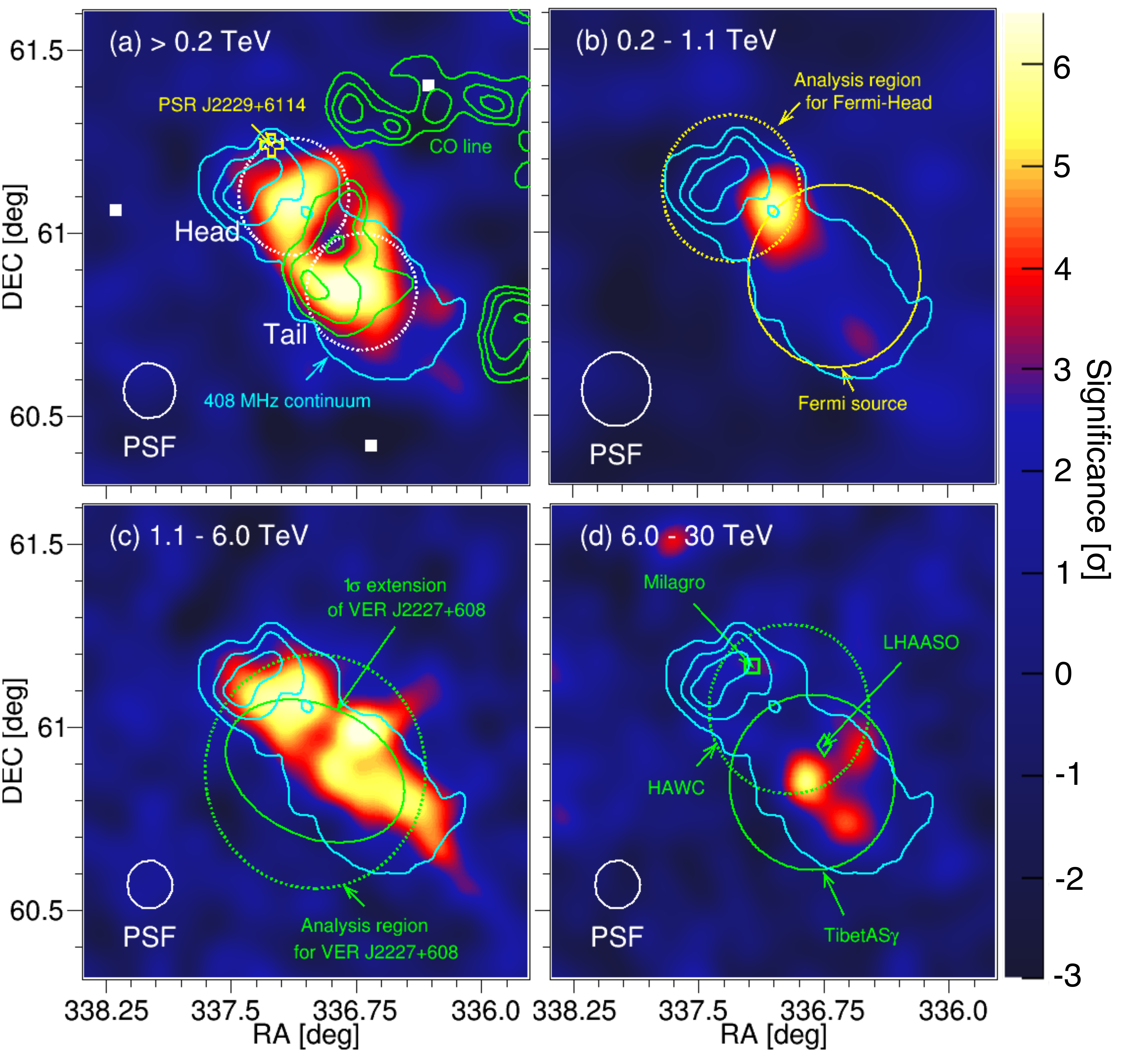}
  \caption{
  Energy-dependent pre-trial significance maps of SNR G106.3+2.7 observed with the MAGIC telescopes.
  {\bf (a)} The map above 0.2~TeV. 
  The white circle labeled "PSF" represents the 0.075$^{\circ}$ size of a Gaussian kernel (corresponding to the MAGIC $\gamma$-ray point spread function) for this analysis.
  The position of PSR J2229.0+6114 is marked with the open yellow cross.
  The cyan contours (overlaid on all panels) show the radio emission of SNR G106.3+2.7 at 408 MHz by DRAO~\citep{Pineault_2000}.
  The green contours represent $^{12}$CO ($J=1-0$) line intensity integrated over the velocity range from $-6.41$ to $-3.94$~km~s$^{-1}$.
  The white dotted circles show $\theta^{2}$ cut regions of the {\it head} and {\it tail} regions, respectively, as shown in Table~\ref{tab:SourceModel}. 
  Also shown by white squares are the pointing positions used in the observations.
  {\bf (b)} The map at 0.2--1.1~TeV.
  The white circle labeled "PSF" represents the 0.100$^{\circ}$ size of a Gaussian kernel as the panel (a). The yellow solid and dotted circle represent the extension and location of the \textit{Fermi}-LAT source~\citep{Xin:2019xeb} and the analysis region for the {\it head} region used in~\cite{Liu:2020gss}, respectively.
  {\bf (c)} The map at 1.1--6.0~TeV. 
  The white circle labeled "PSF" represents the 0.065$^{\circ}$ size of a Gaussian kernel as the panel (b). The green ellipse and dotted circle represent the extended TeV $\gamma$-ray emission of VER J2227+608 and $\theta^{2}$ cut region used in the VERITAS paper~\citep{Acciari:2009zz}, respectively.
  {\bf (d)} The map at 6.0--30~TeV. The white circle labeled "PSF" represents the 0.065$^{\circ}$ size of a Gaussian kernel as the panel (c). The green solid and dotted circle represent the extended $\gamma$-ray emission above 10~TeV observed with Tibet AS$\gamma$~\citep{Tibet2021NatAs} and the upper limit at $90\%$ confidence level of the Gaussian extension of HAWC J2227+610~\citep{Albert:2020ngw}.
  The open square and diamond show the centroid of the VHE $\gamma$-ray emission detected with Milagro~\citep{Abdo:2009ku} and LHAASO~\citep{Cao2021}, respectively.}
  \label{fig:MAGIC_SkyMap}
\end{figure*}

The pre-trial significance maps around VER J2227+608/SNR G106.3+2.7 in different energy bands are shown in Fig.~\ref{fig:MAGIC_SkyMap}.
The panel (a) of the figure shows 
the morphology of the $\gamma$-ray emission above 0.2~TeV
overlaid with the radio emission contours at 408MHz
measured by DRAO~\citep{Pineault_2000} and $^{12}$CO ($J=1-0$) emission contours~\citep{Taylor_2003}.
$\gamma$-ray emission above 0.2~TeV from the direction of VER 2227+608 is clearly detected.
Integrating the same area as VERITAS and using Eq.~17 of~\cite{Li:1983fv}, the statistical significance is $8.9\sigma$. 
It is extended and spatially coincident with the radio shell of the SNR, i.e., the emission region is extending from the SNR {\it head} region to the {\it tail} region.
The emission at the {\it tail} coincides with strong $^{12}$CO ($J=1-0$) emission, but the overall emission profile does not follow well the CO distribution. The emission at the {\it head} is in fact seen where $^{12}$CO ($J=1-0$) emission is not observed. It should be noted that $^{12}$CO ($J=1-0$) does not trace all existing interstellar gas as will be discussed in Sect.~\ref{section:Discussion}.

The panels (b), (c) and (d) of Fig.~\ref{fig:MAGIC_SkyMap} show the maps at 0.2 to 1.1~TeV, 1.1 to 6.0~TeV and 6.0 to 30~TeV, respectively. 
The morphology of the detected $\gamma$-ray emission clearly changes with energy.
By fitting with a symmetric Gaussian function, the center position of the $\gamma$-ray emission in the highest energy band of 6.0--30\,TeV is estimated to be (RA, DEC) = ($336.66 \pm 0.05^{\circ}$, $+60.87 \pm 0.02^{\circ}$) (J2000), which is offset from the location of PSR J2229+6114 by $0.47 \pm 0.03^{\circ}$ (Panel d).
On the other hand, the lower energy emission extends close to the pulsar position (Panels b and c). The centroid of the low energy emission for 0.2--1.1~TeV and its distance from the pulsar position are found to be (RA, DEC) = ($336.99 \pm 0.04^{\circ}$, $+61.04 \pm 0.02^{\circ}$) (J2000) and $0.24 \pm 0.03^{\circ}$.
The $1\sigma$ extension at 6.0--30\,TeV after removing the effect of PSF is $0.14\pm0.09^{\circ}$, which is consistent with the value ($0.24\pm0.14^{\circ}$) reported by Tibet AS$\gamma$~\cite{Tibet2021NatAs}.

\begin{figure*}
   \begin{tabular}{cc}
  \begin{minipage}{0.5\hsize}
    \centering
    \includegraphics[width=\hsize]{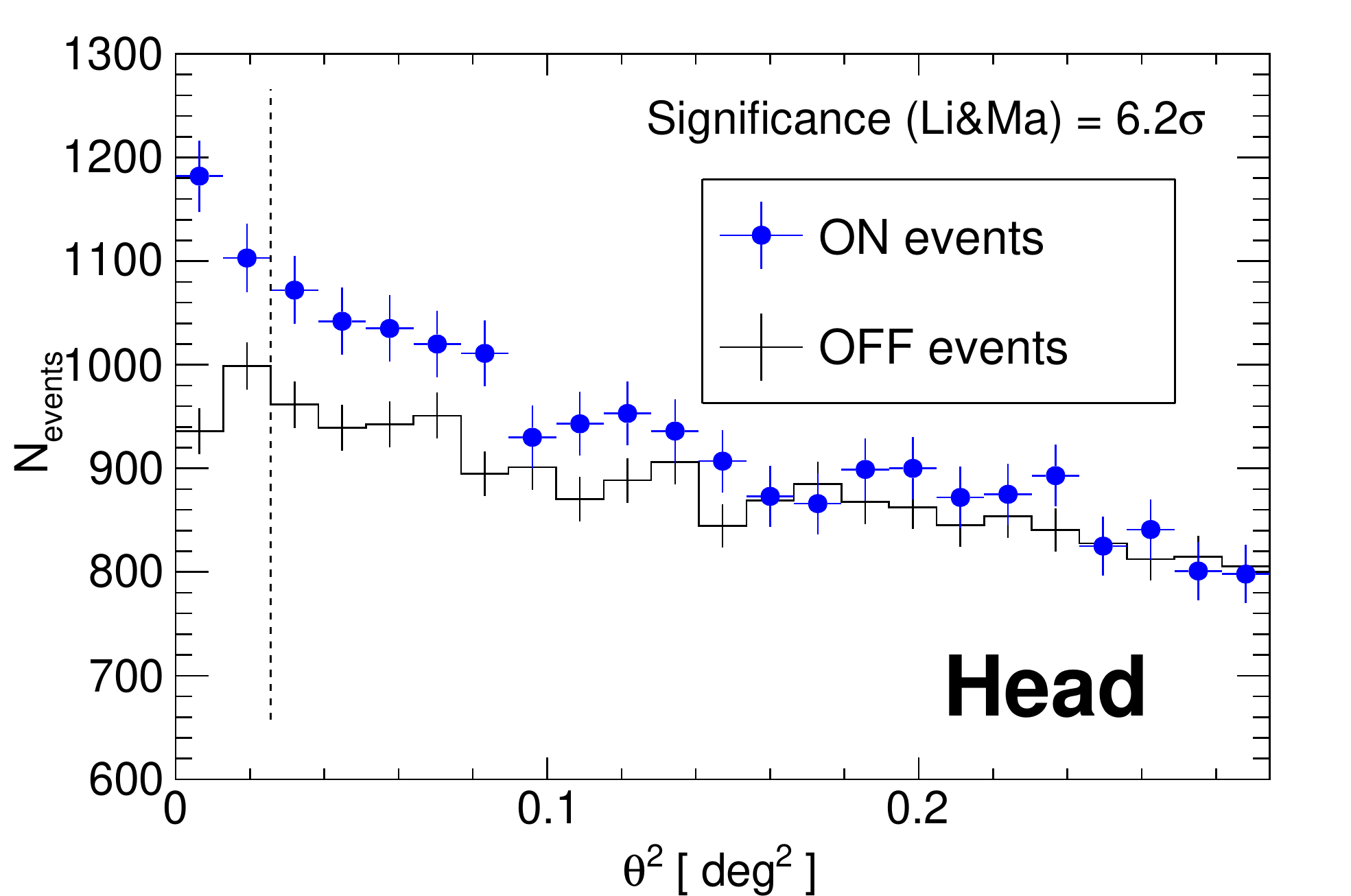}
  \end{minipage} 
  \begin{minipage}{0.5\hsize}
    \centering
    \includegraphics[width=\hsize]{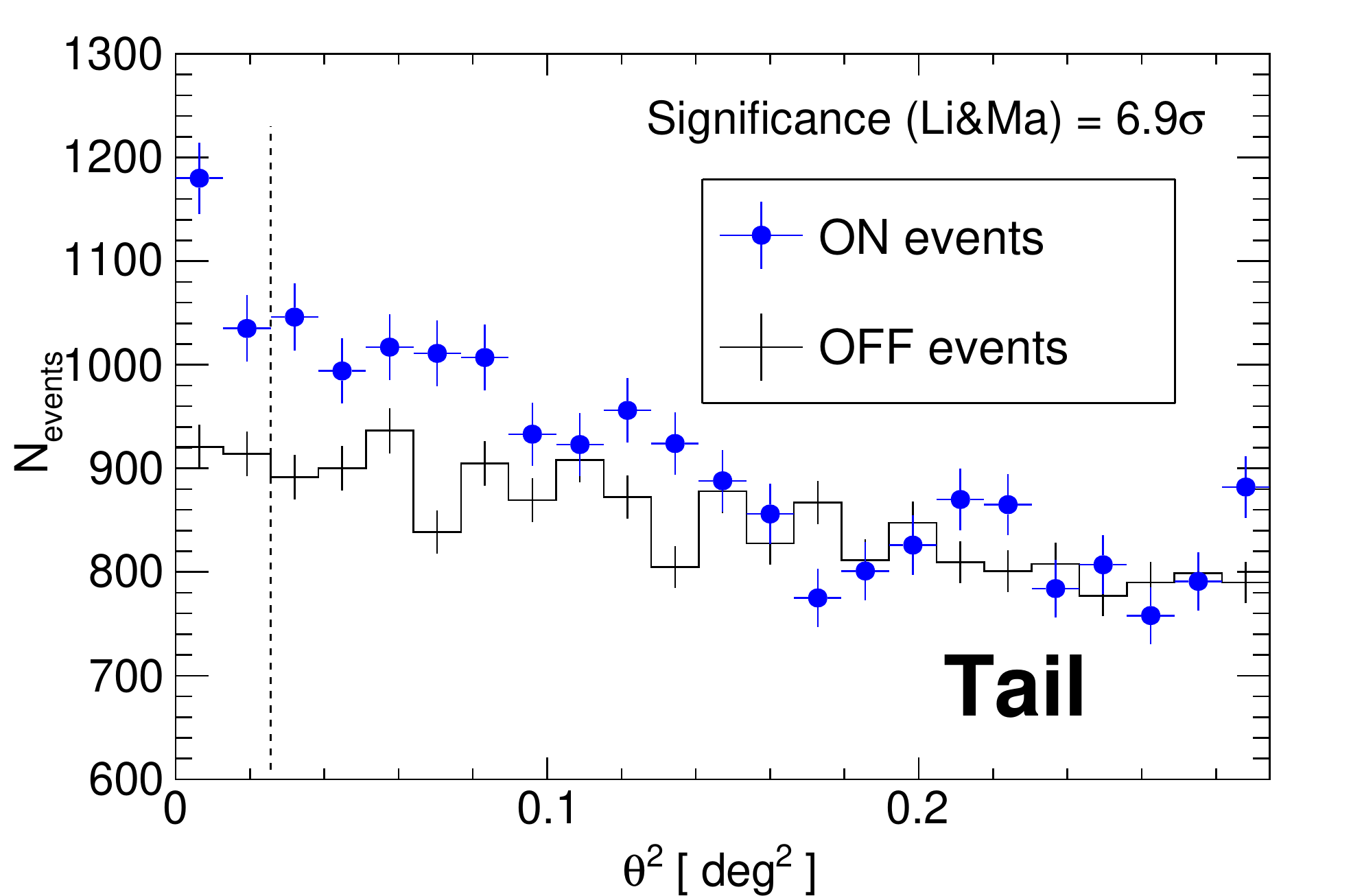}
  \end{minipage}
  \end{tabular}
  \caption{$\theta^{2}$ distributions of ON (blue circles) and OFF (black line) events above 0.2~TeV toward the center of the {\it head} region (left) and that of the {\it tail} region (right).
  The region between zero and the vertical dashed line (at $\theta^{2}$ = 0.0256 deg$^{2}$) has been used to estimate ON and OFF events.
  The OFF data represent the average of six regions rotated by 120 and 240~deg with respect to each wobble center from the ON region.
  }
  \label{fig:ThetaSquarePlots}
\end{figure*}

To understand the emission mechanism better, we studied the $\gamma$-ray spectra at the {\it head} and the {\it tail} regions.
The parameters of the {\it head} and the {\it tail} regions are summarized in Table~\ref{tab:SourceModel} and shown in Fig.~\ref{fig:MAGIC_SkyMap} (a).
The centers of these regions are obtained from a fit to the $\gamma$-ray map above 0.2~TeV (Fig.~\ref{fig:MAGIC_SkyMap} (a)) with a double symmetric Gaussian.
The position of the {\it tail} emission is in good agreement with the peak position observed with VERITAS/Tibet~\citep{Acciari:2009zz, Tibet2021NatAs} and included within the upper limit at $90\%$ confidence level of the Gaussian extension of HAWC J2227+610~\citep{Albert:2020ngw}.
The spatial distribution in Fig.~\ref{fig:MAGIC_SkyMap}(a) appears to have a more complex shape than the double symmetric Gaussian function, but as discussed in Appendix~\ref{section:residual}, current statistics allow fitting data with this function.
The radii of these areas are chosen to be the same for both regions and of maximum length without the regions overlapping.
In Fig.~\ref{fig:ThetaSquarePlots}, we show the so-called $\theta^{2}$ distributions of the two regions, where $\theta$ is the opening angle between the center of the region and the event arrival direction.
For each of the three wobble-pointing positions, two OFF regions were defined such that the ON and the two OFF regions form an equilateral triangle with its center at the camera center.
The OFF events are estimated by taking the average of these six regions.
The excesses are detected from the {\it head} and {\it tail} regions above 0.2~TeV with statistical significance of $6.2\sigma$ and $6.9\sigma$, respectively, evaluated using Eq.~17 of~\cite{Li:1983fv}.
The significances for 0.2--1.1~TeV are $4.8\sigma$ at {\it head} and $2.8\sigma$ at {\it tail}, while for 
6.0--30\,TeV they are $6.5\sigma$ at the {\it tail}, and only $2.4\sigma$ at the {\it head}, indicating that the magnitude ratio of the {\it head} and the {\it tail} emissions flips between the low and high energy bands.

\begin{table}
  \centering
  \caption{Regions considered in this work for the analysis of MAGIC data and their modelling.}
    \begin{tabular}{lcccc} \\ \hline \hline
      Source          & RA             & DEC               & Radius \\ \hline
      {\it head} region & $337.^\circ13$  &  $61.^\circ10$    & $0.^\circ16$ \\
      {\it tail} region & $336.^\circ72$  &  $60.^\circ84$    & $0.^\circ16$ \\ \hline
    \end{tabular}
    \label{tab:SourceModel}
\end{table}

\begin{figure}
\centering
\includegraphics[width=\hsize]{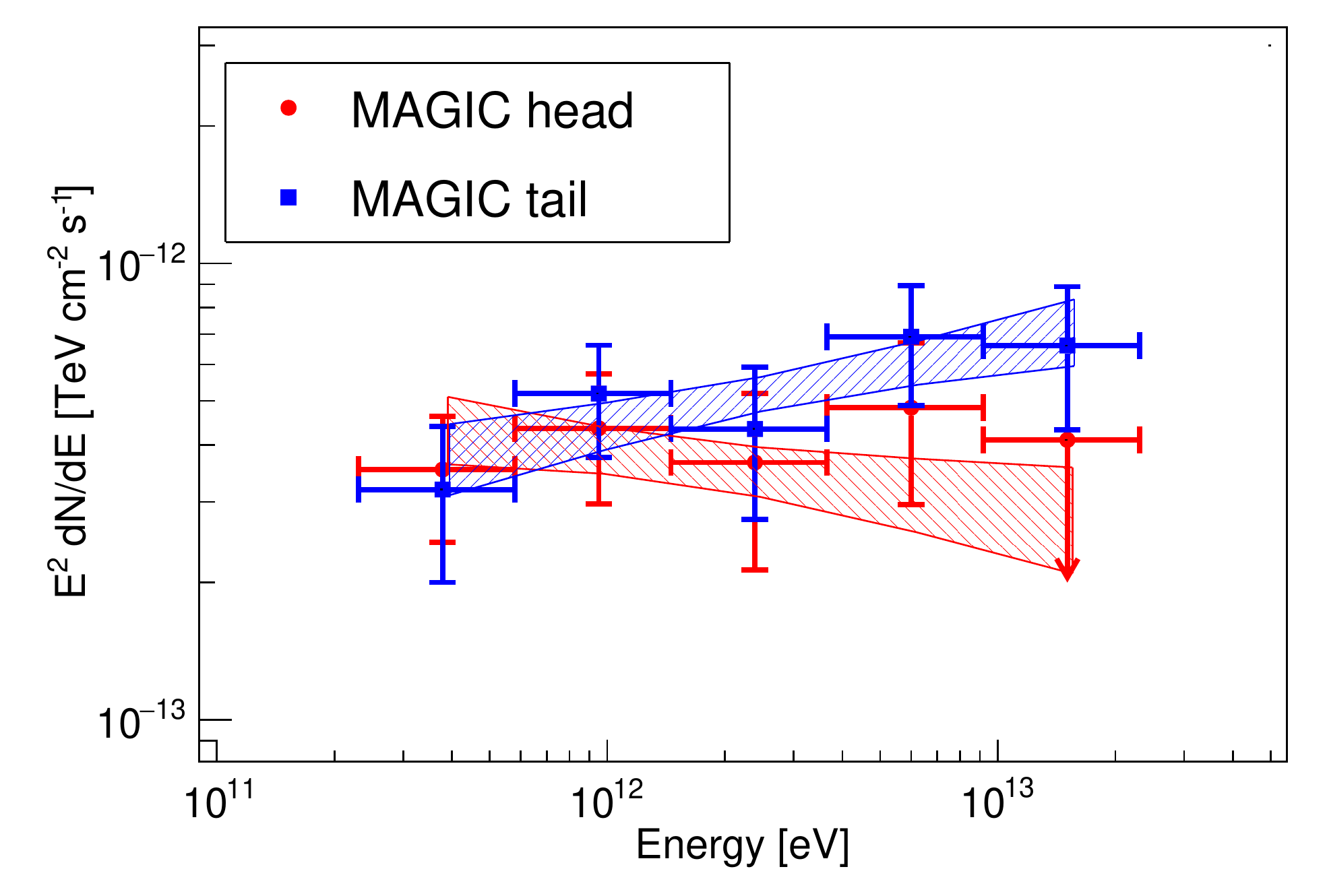}
\caption{Energy spectra of the {\it head} and {\it tail} regions.
Red and blue data represent the spectra of the {\it head} and {\it tail}, respectively.
The color bow-tie areas show the result of fitting with a simple power-law function and $1\sigma$ statistical uncertainties.
}
\label{fig:HeadTailSED}
\end{figure}

\begin{figure*}
\centering
\includegraphics[width=\hsize]{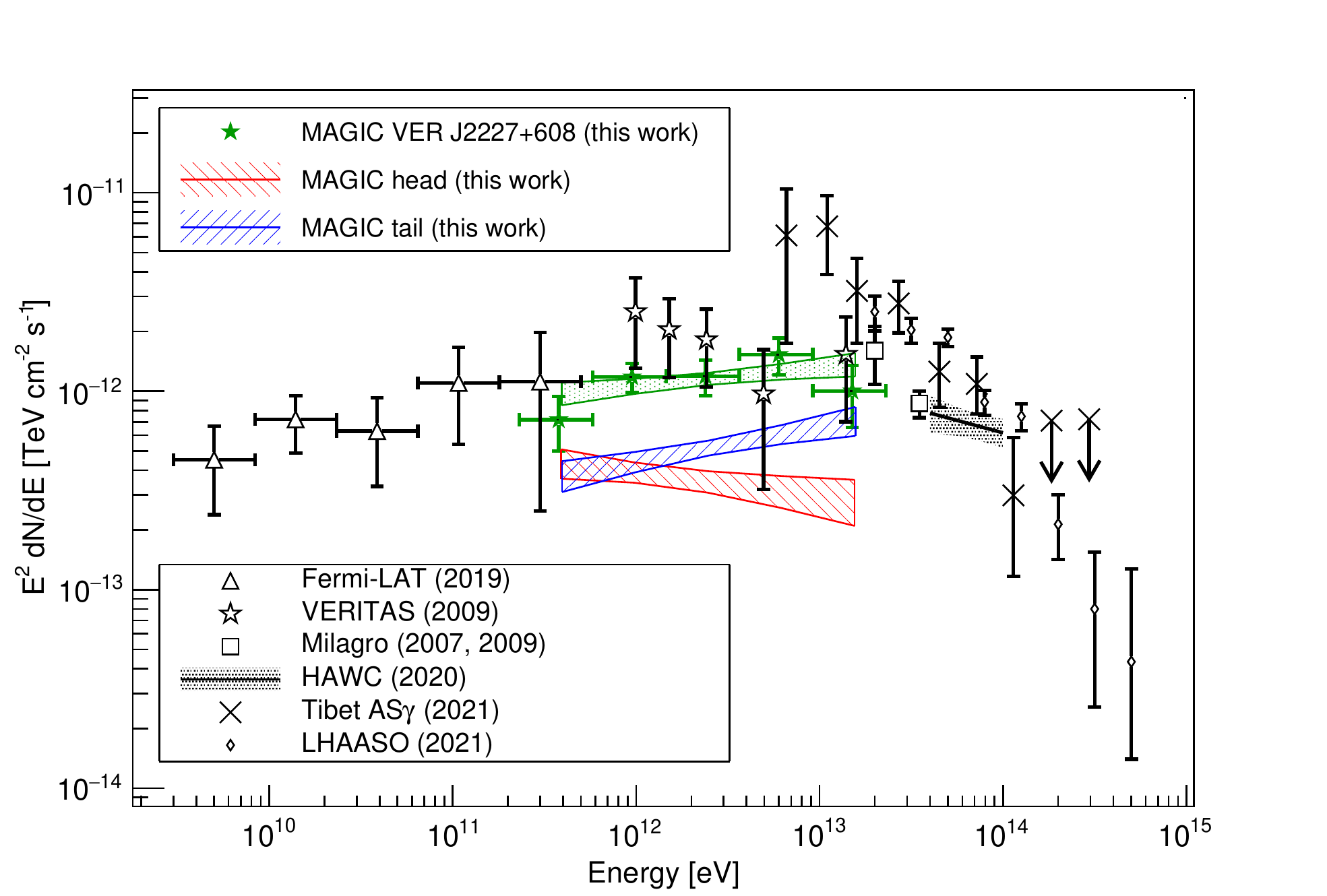}
\caption{
Spectral energy distribution of the whole region of SNR G106.3+2.7.
Green data represents the spectrum of VER J2227+608 region as measured with the MAGIC telescopes.
Also shown as the shaded blue and red regions are the same as in Fig.~\ref{fig:HeadTailSED}.
The open triangles, open stars, open squares, x marks, and open diamonds show the \textit{Fermi}-LAT~\citep{Xin:2019xeb}, VERITAS~\citep{Acciari:2009zz}, Milagro~\cite{Abdo:2007ad, Abdo:2009ku}, Tibet AS$\gamma$~\citep{Tibet2021NatAs}, and LHAASO measurements~\citep{Cao2021}, respectively. 
The black bow-tie area shows a power-law fit and $1\sigma$ statistical errors measured by HAWC~\citep{Albert:2020ngw}.
}
\label{fig:GeVTeVSED}
\end{figure*}

Fig.~\ref{fig:HeadTailSED} and \ref{fig:GeVTeVSED} show the $\gamma$-ray spectra of the two regions defined in Table~\ref{tab:SourceModel} and the extraction region of VER J2227+608~\citep{Acciari:2009zz}, respectively.
Using the forward-folding method~\citep{Aleksic:2014lkm}, the spectra are fitted with a power-law function:
\begin{equation}
  \label{equ:powerlaw}
  \frac{dN}{dE} = N_{0} \left( \frac{E}{3~\mathrm{TeV}} \right)^{-\Gamma}.
\end{equation}
The best-fit parameters are summarized in Table~\ref{tab:Fitting}.
The $\gamma$-ray spectrum in the {\it tail} region has a higher flux and a marginally harder index than that of the {\it head} region.
For the VER J2227+608, using the same integration region as VERITAS, our results are consistent with theirs \citep{Acciari:2009zz} within the statistical uncertainties in both the index and the normalization at 3~TeV.
The apparent discrepancy seen in Fig.~\ref{fig:GeVTeVSED} between the MAGIC results and the Tibet AS$\gamma$ measurement at the 6--20\,TeV range, amounts to only $1.4\sigma$ statistical uncertainty.
Considering the source extension of VER J2227+608 and the MAGIC PSF, the flux derived in this work may correspond to $\sim60\%$ of the whole region estimated with the other experiments.
If this loss is considered, the discrepancy between MAGIC and Tibet AS$\gamma$ relaxes from $1.4\sigma$ to $1.1\sigma$. In addition, if the systematic uncertainties are taken into account, both results agree within $1\sigma$.


\begin{table*}
\begin{threeparttable}
  \centering
    \caption{Comparison of the spectral parameters between this MAGIC results reported here and the VERITAS ones~\citep{Acciari:2009zz}. All sources were fitted with the power-law function of Eq.~\ref{equ:powerlaw}, using a forward-folding method~\citep{Aleksic:2014lkm}.}
    \begin{tabular}{cccc} \\ \hline \hline
      Source  & $N_{0}$ ($10^{-14}$ cm$^{-2}$ s$^{-1}$ TeV$^{-1}$) at 3\,TeV & $\Gamma$ & $\chi^{2}$/ndf \\ \hline 
      {\it head}  & 3.8 $\pm$ 0.7$_{\mathrm{stat}}$ $\pm$ 0.7$_{\mathrm{sys}}$ &  2.12 $\pm$ 0.12$_{\mathrm{stat}}$ $\pm$ 0.15$_{\mathrm{sys}}$  & 5.5/6 \\ 
      {\it tail} & 6.0 $\pm$ 0.7$_{\mathrm{stat}}$ $\pm$ 1.0$_{\mathrm{sys}}$ &  1.83 $\pm$ 0.10$_{\mathrm{stat}}$ $\pm$ 0.15$_{\mathrm{sys}}$ & 2.6/6 \\
      VER J2227+608 (MAGIC) & 13.1 $\pm$ 1.1$_{\mathrm{stat}}$ $\pm$ 2.1$_{\mathrm{sys}}$ & 1.91 $\pm$ 0.07$_{\mathrm{stat}}$ $\pm$ 0.15$_{\mathrm{sys}}$ & 7.1/6 \\
       VER J2227+608~\citep[VERITAS,][]{Acciari:2009zz}  & 11.5 $\pm$ 2.7$_{\mathrm{stat}}$ $\pm$ 3.5$_{\mathrm{sys}}$ &  2.3  $\pm$ 0.33$_{\mathrm{stat}}$ $\pm$ 0.30$_{\mathrm{sys}}$ & - \\ \hline
    \end{tabular}
    \label{tab:Fitting}
\end{threeparttable}
\end{table*}

%--------------------------------------------------------------------
\section{Modelling} \label{section:Modelling}
Previous studies~\citep[e.g.,][]{Liu:2020gss, Ge:2020uft, Bao:2021xdf, Fang:2022uge} discussed the origin of $\gamma$ rays using the spectrum up to 100~TeV of the whole region of this object, while the $\gamma$-ray spectra of the {\it head} and the {\it tail} regions are obtained in this work for the first time.
Here, we try to model the $\gamma$-ray emission mechanism of the {\it head} and the {\it tail} region individually. 
Both hadronic and leptonic models are examined using the naima framework~\citep{naima}.

\subsection{Description of VHE $\gamma$-ray emission}
The spatial coincidence of the MAGIC VHE $\gamma$-ray emission and the 408~MHz radio continuum shown in Fig. \ref{fig:MAGIC_SkyMap} (a) suggests that the VHE $\gamma$-ray emission is associated with the radio SNR G106.3+2.7.
On the other hand, as shown in Fig.~\ref{fig:MAGIC_SkyMap}(d), the significant $\gamma$-ray emission above 6.0~TeV is detected in the {\it tail} region but not in the {\it head} region. 
The extracted spectra, shown in Fig.~\ref{fig:GeVTeVSED}, suggest that the {\it head} contribution to the total flux above 10~TeV is less than $37.1\%$ ($2\sigma$ upper limit). 
In the following modelling and discussion, we assume that the measured emission above 10~TeV~\citep{Abdo:2009ku, Albert:2020ngw, Tibet2021NatAs, Cao2021} is only from the {\it tail} region.

\subsection{SNR G106.3+2.7 and measurements in other wavelengths}
The distance to the SNR G106.3+2.7 from the Earth is assumed to be 0.8~kpc~\citep{Kothes_2001}\footnote{Once we assume that the distance is 3~kpc estimated from the X-ray observation~\citep{Halpern:2000qh} instead of 0.8~kpc, the estimate of SNR size is $(3/0.8~\rm{kpc}) \sim 4$ times larger and also the total energy of particles ($W$) in the modelling is $(3/0.8~\rm{kpc})^{2} \sim 14$ times higher. However, these do not affect the results discussed in the text.}. 
\cite{Pineault_2000} derived the radio fluxes from the SNR-{\it head} and {\it tail}, separately. 
We adopted them since the definition of {\it head} and {\it tail} are (not perfectly but) nearly identical between this work and \cite{Pineault_2000}.
The X-ray spectra for the {\it head} and {\it tail} regions are taken from results of the "East" and "West" regions from \cite{Fujita:2021fzh}, respectively, multiplying the intensity by the area of a circle with a radius of 0.16$^\circ$ used in the MAGIC analysis.
At GeV range, \cite{Xin:2019xeb} and \cite{Liu:2020gss} reported the spectral points and upper limits assuming that the sources have a disk shape. 
They obtained the radii of 0.20$^\circ$ and 0.25$^\circ$ for the disks.
We scaled down their measurements by $(0.16/0.20)^{2}$ for the {\it head} and $(0.16/0.25)^{2}$ for the {\it tail.
}
In this study, we do not consider the direct contributions from the compact Boomerang nebula, whose angular diameter is $\sim$ 0.05$^\circ$, because the $\gamma$-ray flux of the region is estimated to be $\sim10\%$ or less of the {\it head} region from the radio and X-ray flux~\citep{Liu:2020gss}.

\subsection{Leptonic Model} \label{sec:leptonicmodel}
For the leptonic model, the VHE $\gamma$-ray emission can be mainly produced by inverse Compton (IC) scattering~\citep{Blumenthal1970RvMP}.
The energy spectra of electrons are assumed to follow a power-law function with an exponential cutoff.
The cosmic microwave background, a galactic near-infrared (NIR) radiation field, and a galactic far-infrared (FIR) radiation field are considered as seed photon fields in the IC process.
Using the model included in the GALPROP package~\citep{Porter:2008ve}, the energy density of NIR and FIR are estimated to be 0.1~eV~cm$^{-3}$ at $T$ = 30~K and 0.3~eV~cm$^{-3}$ at $T$ = 3000~K, respectively.
The radio and the non-thermal X-ray emission are produced by high-energy electrons via the synchrotron process.

The following procedure obtained the model parameters: the total amount of electrons is determined to reproduce the $\gamma$-ray data with the given target photon density described above, and the magnetic field strength and electron cutoff energy are determined such that the synchrotron reproduces the radio and X-ray data, respectively.

\subsection{Hadronic Model}
For the hadronic model, the $\gamma$-ray emission results from the decay of neutral pions produced by inelastic pp-collisions.
The energy spectra of protons are assumed to follow a power-law function with an exponential cutoff.
The target gas density of each region is estimated using the radio line data of HI and $^{12}$CO ($J=1-0$) (see Appendix~\ref{section:gasdensity}).
As a result, we adopted $n_{\mathrm{HI}} + n_{\mathrm{CO}} \sim 100\,\mathrm{cm^{-3}}$ for the {\it head} region and $n_{\mathrm{HI}} + n_{\mathrm{CO}} \sim 200\,\mathrm{cm^{-3}}$ for the {\it tail} region.
Furthermore, IC and synchrotron emissions by relativistic electrons are also considered as in Sect.~\ref{sec:leptonicmodel}.

The proton spectrum (flux and energy cutoff) is determined to reproduce the $\gamma$-ray data, while the electron spectrum is given such that the synchrotron radiation reproduces the radio and X-ray data assuming a magnetic-field strength of 10~$\rm{\mu G}$.

%---------------------
\begin{figure*}
    \begin{tabular}{cc}
     \begin{minipage}{0.475\hsize}
            \centering
            \includegraphics[width=\hsize]{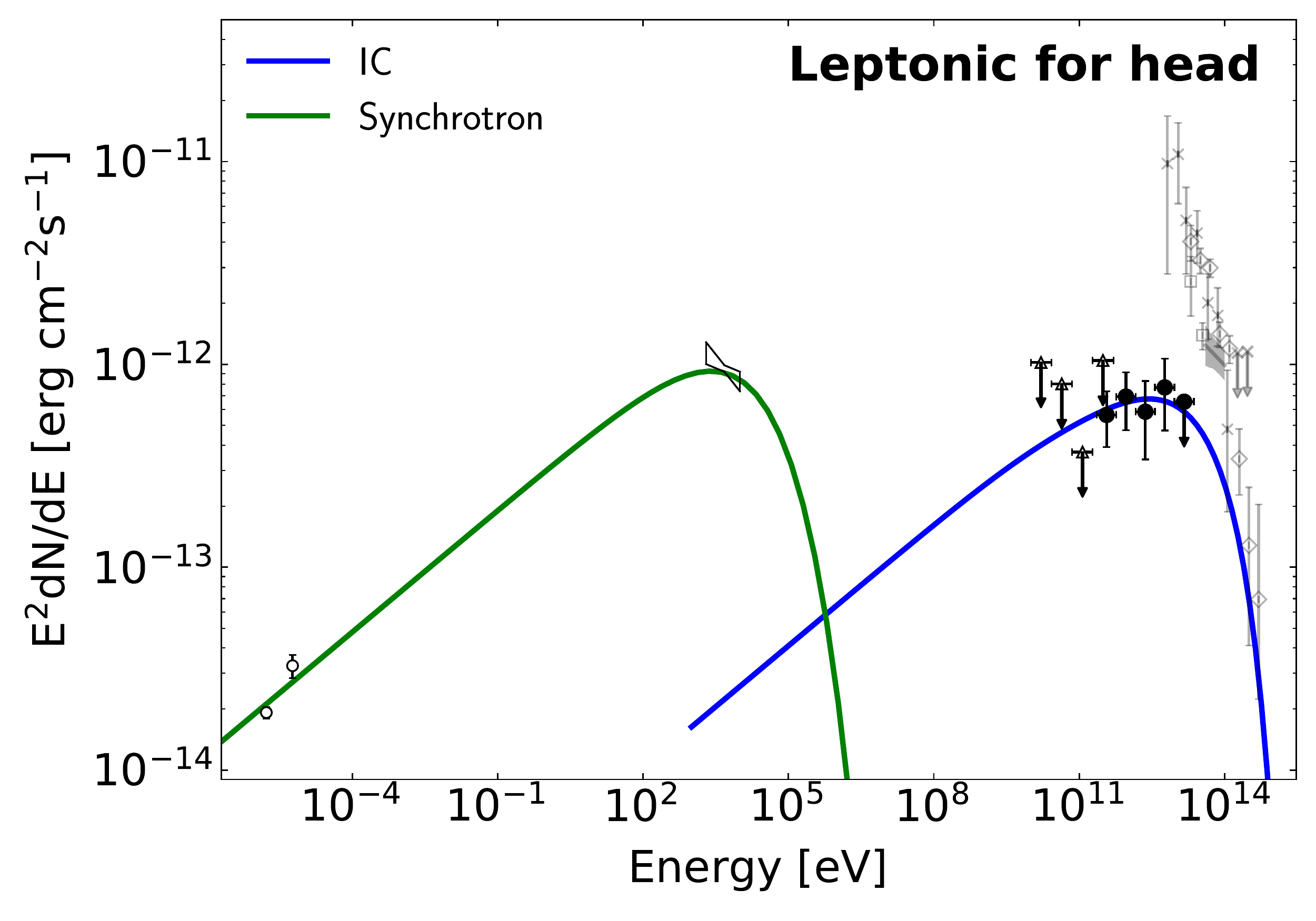}
        \end{minipage}
        &
        \begin{minipage}{0.475\hsize}
            \centering
            \includegraphics[width=\hsize]{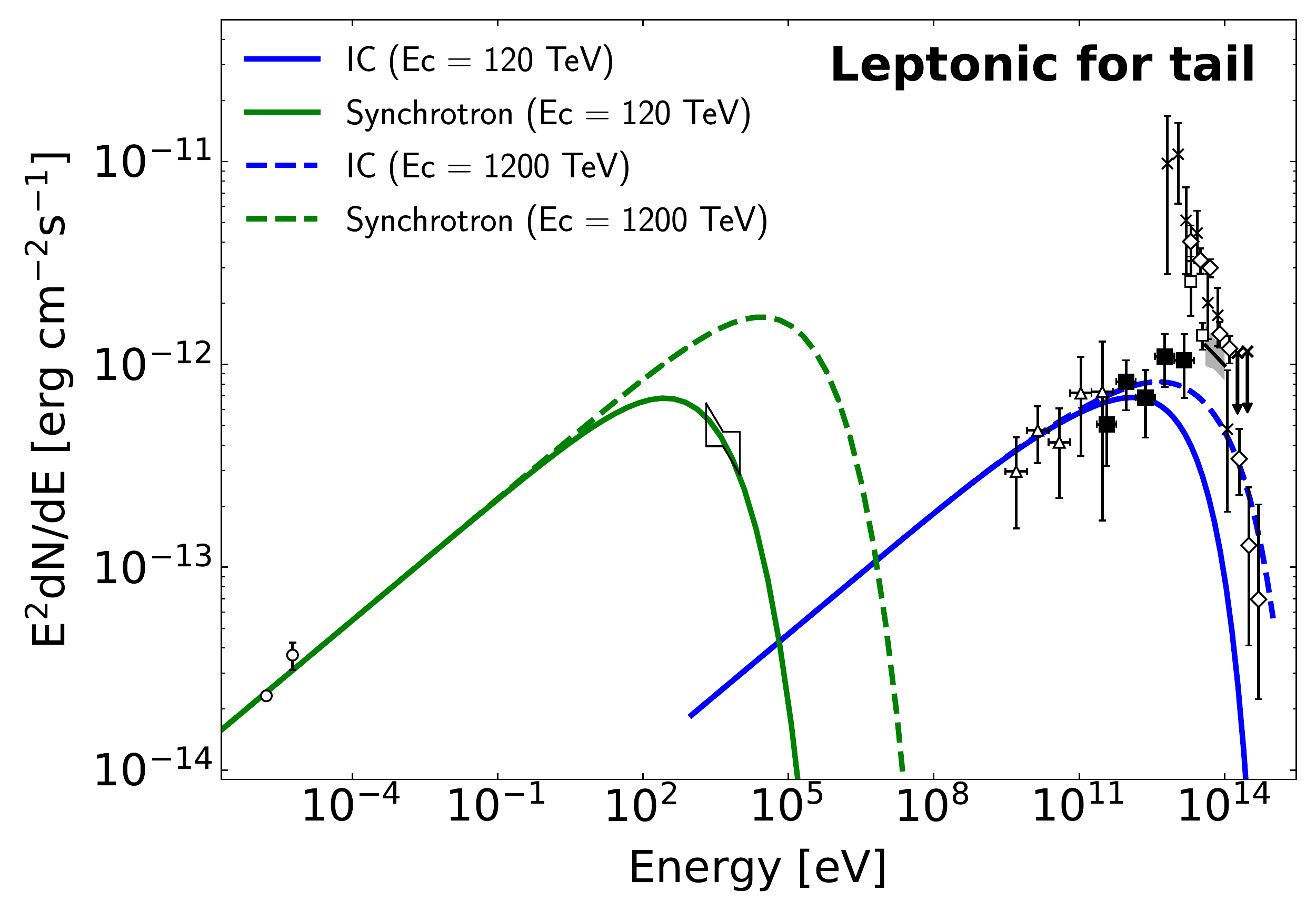}
        \end{minipage}
    \\
    \\
        \begin{minipage}{0.475\hsize}
            \centering
            \includegraphics[width=\hsize]{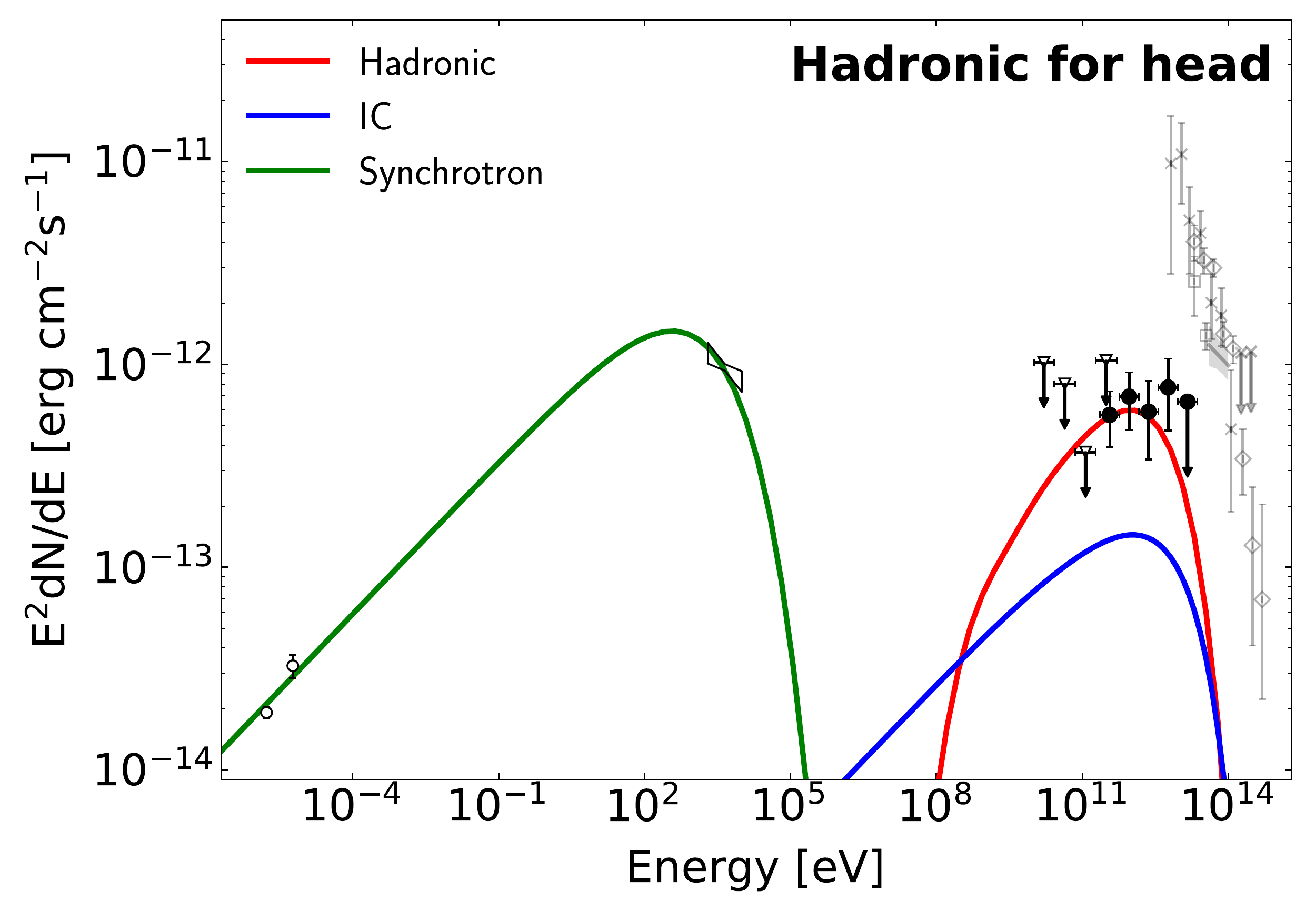}
        \end{minipage}
        &
        \begin{minipage}{0.475\hsize}
            \centering
            \includegraphics[width=\hsize]{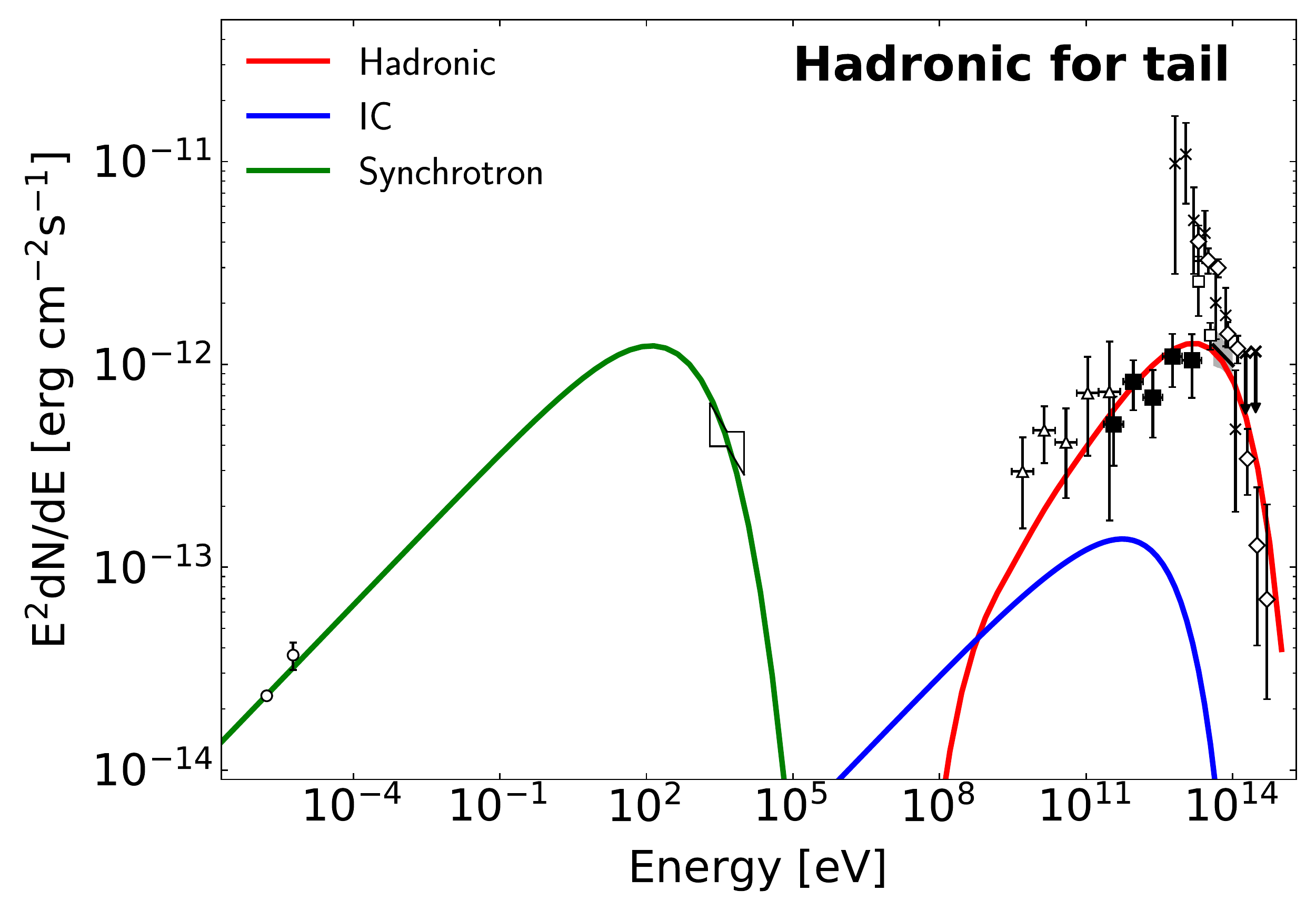}
        \end{minipage}
        \\
  \end{tabular}  
\caption{
Modelling of the SED of SNR G106.3+2.7. 
The left and right panels show the results of the {\it head} and the {\it tail}, respectively.
The top and bottom panels show the results of the leptonic and hadronic models, respectively.
The white circles show the radio flux of each region~\citep{Pineault_2000}.
The black bow-tie area shows a power-law fit and $1\sigma$ statistical errors measured by Suzaku~\citep{Fujita:2021fzh}.
The open triangles show the \textit{Fermi}-LAT measurements~\citep{Xin:2019xeb, Liu:2020gss}.
The markers in the TeV $\gamma$-ray band are the same as in Fig.~\ref{fig:GeVTeVSED}, but those corresponding to the MAGIC data are shown in black here.
The red, blue and green lines show the hadronic, IC and synchrotron emission.
The model parameters of each panel are summarized in Table~\ref{tab:Modeling}.
In the top-right panel, the solid and dashed lines show the leptonic emission with the energy cutoff of 120 and 1200~TeV, respectively.
}
\label{fig:Modeling}
\end{figure*}

\begin{table*} 
\begin{threeparttable}
  \centering
    \caption{
    Model parameters for reproducing the observed spectra.
    $\alpha$ and $E_{\mathrm{c}}$ are the power-law index and the cutoff energy of the particle spectrum, respectively.
    $W$ is the total energy of particles with energy above 1~GeV. The subscript e and p denote electrons and protons.
    $B$ is the magnetic-field strength in unit of $\rm \mu$G.
    $N_{\rm gas}$ is the target gas density in unit of ${\rm cm^{-3}}$.
    }
    \label{tab:Modeling}
    \begin{tabular}{c|c|c|c|c|c|c|c|c|c|c} \hline \hline
    Model & Source & $\alpha_{\rm e}$ & $E_{\rm c, e}$ [TeV] & $W_{\mathrm{e}}$ [erg] & $B$ [$\rm \mu$G] & $\alpha_{\rm p}$ & $E_{\rm c, p}$ [TeV] & $W_{\mathrm{p}}$ [erg] & $N_{\rm gas} {\rm[cm^{-3}]}$ & $\chi^{2}$/ndf \\ \hline 

    \multirow{2}{*}{Leptonic} & {\it head} & 2.6 & 360 & 1.4 $\times$ 10$^{47}$ & 3 & - & - & - & - & 5.0/7 \\
    & {\it tail} & 2.6 & 120 (1200)\tnote{$\dagger$} & 1.6 $\times$ 10$^{47}$ & 3 & - & - & - & - & 103.1/31 ($\gg1$)\tnote{$\dagger$} \\ \hline
    
    \multirow{2}{*}{Hadronic} & {\it head} & 2.5 & 60 & 1.8 $\times$ 10$^{46}$ & 10 & 1.7 & 60 & 8.9 $\times$ 10$^{45}$ & 100 & 5.3/7 \\
    & {\it tail} & 2.5 & 35 & 2.0 $\times$ 10$^{46}$ & 10 & 1.7 & 1000 & 8.2 $\times$ 10$^{45}$ & 200 & 39.9/31 \\ \hline
    \end{tabular}
    \begin{tablenotes}
     \item[$\dagger$] In the top-right panel of Fig.~\ref{fig:Modeling}, the model curve using the value in the parentheses is shown with the dashed line.
    \end{tablenotes}
\end{threeparttable}
\end{table*}

\subsection{Results of modelling}
The Fig.~\ref{fig:Modeling} show the modelling result of the leptonic (upper panels) and hadronic (lower panels) models.
Parameters for the modelling are summarized in Table~\ref{tab:Modeling}.

The broad-band spectrum of the {\it head} region can be explained well with the leptonic model ($\chi^{2}$/ndf = 5.0/7\footnote{
Since the results of the X-ray band~\citep{Fujita:2021fzh} and HAWC~\citep{Albert:2020ngw} have been given as a fitted power-law function, the flux and statistical uncertainty only at the normalization energy of the fit are considered in the calculation of the chi-squared statistic.
In addition, these calculations for $\gamma$-ray observation data take into account not only statistical errors but also systematic errors on the normalization flux. As the systematic errors for the \textit{Fermi}-LAT and LHAASO results were not estimated in the previous papers for this source, we estimate those of \textit{Fermi}-LAT and LHAASO with the uncertainties of the effective area (\url{https://fermi.gsfc.nasa.gov/ssc/data/analysis/scitools/Aeff_Systematics.html}) and absolute energy scale~\citep{LHAASO:2021xzn}, respectively.
}). 
In the case of the {\it tail} region, the leptonic model can reproduce the observed data only in the radio, X-ray, \textit{Fermi}-LAT and MAGIC band ($\chi^{2}$/ndf = 8.2/13), but fails when including air-shower experiments ($\chi^{2}$/ndf = 103.1/31).
To explain the $\gamma$-ray emission above 10~TeV measured by air shower experiments, a high cutoff energy of electrons of $\sim$1200~TeV is required. 
However, the synchrotron spectrum produced with such high cutoff energy is excluded by the observed X-ray flux.
The $\chi^{2}$/ndf for the model with the high cutoff energy is found to be $\gg$~1 when considering the X-ray data.

For the hadronic model, the $\gamma$-ray spectra of both the {\it head} and the {\it tail} region can be reproduced assuming a proton maximum energy of 60~TeV and 1~PeV, respectively ($\chi^{2}$/ndf = 5.3/7 and 39.9/31). 
While the $\gamma$-ray emission has a hadronic origin, the observed data in the radio and X-ray band may instead result from synchrotron emission.
The parent electron distribution should follow a power-law spectrum different from the ones of protons (parameters shown in Table~\ref{tab:Modeling}).

\section{Discussion} \label{section:Discussion}
\subsection{{\it Head} region} \label{sec:headmodel}
The X-ray emission in the {\it head} region exhibits a softening of the spectral index with distance from the pulsar, suggesting the emission originates in electrons accelerated in and propagated from the shock of the PWN~\citep{Ge:2020uft}.
Our modelling result shows that X-ray and gamma-ray fluxes can be explained with leptonic emission from the same electron population. It thus implies that the gamma-ray emission can originate in the PWN.
Assuming the electron cutoff energy ($E_{\rm{c, e}}$) and the magnetic-field strength ($B$) used in the leptonic model for the {\it head} region, the electron lifetime due to synchrotron losses is given by: $\sim$3.9~kyr $(E_{\rm{c, e}}/360~\rm{TeV})^{-1}(B/3~\rm{\mu G})^{-2}$, which is consistent with the age of the SNR estimated to be 3.9~kyr from the spectral break in the radio spectrum of the PWN~\citep{2006ApJ...638..225K} or 10~kyr from the pulsar spin-down age~\citep{Halpern:2001fc}.

Hadronic scenario also works for the head. 
The protons accelerated up to 60~TeV can explain the VHE $\gamma$-ray emission detected by MAGIC, given the presence of dense HI clouds in the {\it head} region pointed out by \citet{Kothes_2001}.
Although CO emission is not prominent, HI/CO intensity suggests the presence of gases with a total proton density of $\sim100~{\rm cm}^{-3}$, which is sufficient for the pp emission, as derived in Appendix~\ref{section:gasdensity}. 
Still electrons with a largely different spectral index are needed to explain the radio and X-ray emission. 
One of the simplest explanation would be that electrons are mainly from the PWN, while the protons were accelerated in the shell.
Acceleration up to 60~TeV by a 3~kyr SNR is possible~\citep{Cardillo:2015zda}, while an electron lifetime due to the synchrotron cooling is $\sim$2.2~kyr $(E_{\rm{c, e}}/60~\rm{TeV})^{-1}(B/10~\rm{\mu G})^{-2}$.

\subsection{{\it Tail} region} \label{sec:tailmodel}
The modelling described in the previous section suggests that it is difficult to explain the {\it tail} emission with the leptonic model. 
On the other hand, the hadronic model worked well;
the $\gamma$-ray spectrum of the {\it tail} region can be reproduced assuming a proton maximum energy of 1~PeV ($\chi^{2}$/ndf = 39.9/31). 
Generally speaking, acceleration up to 1~PeV can only be achieved at the early stages ($<$ 1.0~kyr) of the SNR evolution~\citep[e.g.,][]{Bell:2013kq, Cardillo:2015zda, Cristofari:2021hbc, Cristofari:2022kqv}.
However, as mentioned in Sect.~\ref{sec:headmodel}, the age of this SNR has been estimated to be 3.9~kyr from the spectral break in the radio spectrum of the PWN~\citep{2006ApJ...638..225K} or 10~kyr from the pulsar spin-down age~\citep{Halpern:2001fc}.
This discrepancy in the SNR age can be solved assuming a CR-escape scenario~\citep[e.g.][]{Aharonian:1996aa, Gabici:2007qb}.
In this scenario, protons accelerated up to $\sim$ PeV energies at a young SNR escape from acceleration regions and illuminate nearby clouds, which produce "delayed" $\gamma$-ray emission. 
This scenario can also explain a proton index of 1.7, harder than 2.0 expected from Diffusive Shock Acceleration (DSA)~\citep[e.g.,][]{Bell:1978zc, Blandford:1978ky}.
On the other hand, it requires high density clouds spatially coinciding with the $\gamma$-ray morphology.
Using the CGPS data of HI and $^{12}$CO ($J=1-0$) (see Appendix~\ref{section:gasdensity}), we confirmed a coincidence of the $\gamma$-ray emission with CO line emission in the velocity range $-6.41$ to $-3.94$~km~s$^{-1}$ in the {\it tail} region, which was already pointed out by~\citet{Kothes_2001} and \citet{Acciari:2009zz}.
This supports the CR-escape scenario in the {\it tail} region.
The scenario is consistent with the interpretation given in~\cite{Albert:2020ngw, Fujita:2021fzh, Tibet2021NatAs}.
The authors estimated the diffusion length of CRs using the relation: $l_{\rm diff}=\sqrt{Dt}$, where $D$ is the diffusion coefficient, and $t$ is the diffusion time.
They then found, even assuming a small diffusion coefficient ($D \sim 10^{26}$~cm$^{2}$s$^{-1}$ at GeV), that the diffusion length for CRs with an energy of $\mathcal{O}$(100~TeV) in 5--10~kyr is larger (40--60~pc) than the size of the SNR ($\sim 6$~pc) and thus suggested the CRs are not confined in the SNR. 
A cloud with a radius of a few pc located at 40 -- 60~pc away is a plausible target considering the energetics of the supernova.

Electrons may also escape in the same way as protons but be affected by radiative cooling, which is not considered in the modelling.
However, the change in the spectral index due to the cooling effect is estimated to be at most 0.1--0.4~\citep{Diesing:2019lwm}, suggesting that the difference ($\sim0.8$) between the proton and electron indices cannot be explained even by considering it.
This fact implies that leptonic and hadronic emissions may happen at different locations and thus under different physical conditions. For example, leptonic emission comes from the SNR shell, while hadronic emission comes from the interstellar gas spatially separated from the SNR.
This assumption can allow the unusual ratio of the total energy of CRs ($W_{\rm p}$ $\lesssim$ $W_{\rm e}$) because only the hadronic emission is affected by the propagation effect~\citep{Gabici:2007qb}, and thus only $W_{\rm p}$ decreases.
Note that the electron lifetime due to synchrotron losses is estimated to be $\sim$3.6~kyr $(E_{\rm{c, e}}/35~\rm{TeV})^{-1}(B/10~\rm{\mu G})^{-2}$, which is in good agreement with the SNR age.

The hard proton index found in the TeV band can also be explained with SNR-cloud interactions~\citep{Inoue:2011xs}, as an alternative to the CR-escape scenario. However, the maximum energy of $\gtrsim$ PeV in SNRs older than 1~kyr cannot be explained with this model. Also, the scenario could not explain the differences in the distribution of electrons and protons~\citep{Diesing:2019lwm}, as mentioned above.

\subsection{Remarks on the discussion}

The integrated region of MAGIC-{\it tail} in this analysis may miss a fraction of the $\gamma$-ray emissions observed by air shower experiments.
Using the Gaussian extension at $>$6~TeV derived with $\theta^{2}$ plot around {\it tail}, the event fraction surviving the $\theta^{2}$ cut is estimated to be 74--95$\%$ ($1\sigma$ uncertainties).
We examined the effect on our model fit for the {\it tail} spectrum when using the scaled flux of air shower experiments by 74$\%$.
In the leptonic model, $\chi^{2}$/ndf changed only slightly (from 103.1/31 to 96.3/31), indicating the model is still inconsistent with the observed data.
In the hadronic model, $\chi^{2}$/ndf also changed (from 39.9/31 to 41.3/31), and the model still works.
As a result, these do not affect our conclusion.

It should also be noted that the data points of Milagro, HAWC, TibetAS$\gamma$, and LHAASO, included in the modelling of the {\it tail} spectrum, are from extraction regions which partially include the {\it head}.
Hence, they are potentially contaminated if the {\it head} emits radiation $>10$~TeV.
Even if, for example, half of the emission above 10 TeV is from the {\it head}, it is not possible to explain the tail emission with this rather simple leptonic model. 

Though more complicated leptonic models, such as two electron populations with adjusted magnetic field strengths can explain the tail emission as demonstrated in \cite{Ge:2020uft}, exploring all possible scenarios with currently obtained data is beyond the scope of this paper.
To accurately determine the emission mechanism, it is first necessary to separate the extraction regions at the {\it head} and {\it tail} also for spectral points above 10~TeV.

%--------------------------------------------------------------------
\section{Summary} \label{section:Summary}
We carried out deep $\gamma$-ray observations of SNR G106.3+2.7 with the MAGIC telescopes.
The MAGIC observations revealed a $\gamma$-ray morphology that is spatially coinciding with the radio emission and achieved a significant detection of TeV $\gamma$ rays from the {\it head} and the {\it tail} regions of SNR G106.3$+$2.7 for the first time.
The energy spectra in energy regimes from 0.2~TeV to 20~TeV of the {\it head} and {\it tail} regions can be well described by a simple power-law function of $dN/dE$ $=$ $N_{0}$ ($E$/3\,TeV)$^{-\Gamma}$ with the indices of $\Gamma$ = $2.12 \pm 0.12$ and $1.83 \pm 0.10$, respectively.
The total flux of the two regions is consistent with the VERITAS results within the statistical uncertainty. 
As the emission above 10~TeV is seen only from the {\it tail} region, it is likely that the $\gamma$ rays above 10~TeV detected with the air shower experiments~\citep[e.g.,][]{Abdo:2009ku} are mainly emitted from the SNR {\it tail}.
We investigated the possibilities to explain the emission from the two regions. 
The {\it head} emission can be explained with both a hadronic and a leptonic model.
Under the assumption that the $\gamma$-ray emission above 10\,TeV is only from the {\it tail} region, the leptonic model emission of the {\it tail} region is in contradiction with the X-ray flux.
The proton spectrum with the cutoff at $\sim 1$~PeV could explain the observed spectrum from the {\it tail} region. It may suggest that protons accelerated in the SNR shock in the past escaped from the SNR and interacted with target gas located in front of the SNR along the line of sight. This scenario could also explain the inconsistency between the SNR age and maximum energy of accelerated protons.
By considering complex particle distributions and/or magnetic field environments, the leptonic model may explain the observed spectra~\citep[e.g.,][]{Ge:2020uft}, but it is beyond the scope of this paper.
For a better determination of the VHE $\gamma$-ray origin, it is necessary to observe the $\gamma$ rays emission $>$10\,TeV with a high sensitivity with an angular resolution better than 0.1$^{\circ}$ enough for resolving the two regions and quantitatively evaluate the difference of the cutoff energies between {\it head} and {\it tail}.
For example, with the current MAGIC telescopes, it would require more 
than $\sim$3600~hours to detect 20--200~TeV emission at the tail.
Such observations could be possible with the new generation of $\gamma$-ray observatories, CTA/ASTRI~\citep{CTAConsortium:2012yjm, Lombardi:2021orw}.

%--------------------------------------------------------------------
% acknowledge
\begin{acknowledgements}
We thank Hidetoshi~Sano, Tsuyoshi~Inoue and Yasuo~Fukui for the fruitful discussion.
We would like to thank the Instituto de Astrof\'{\i}sica de Canarias for the excellent working conditions at the Observatorio del Roque de los Muchachos in La Palma. The financial support of the German BMBF, MPG and HGF; the Italian INFN and INAF; the Swiss National Fund SNF; the grants PID2019-104114RB-C31, PID2019-104114RB-C32, PID2019-104114RB-C33, PID2019-105510GB-C31, PID2019-107847RB-C41, PID2019-107847RB-C42, PID2019-107847RB-C44, PID2019-107988GB-C22 funded by MCIN/AEI/ 10.13039/501100011033; the Indian Department of Atomic Energy; the Japanese ICRR, the University of Tokyo, JSPS, and MEXT; the Bulgarian Ministry of Education and Science, National RI Roadmap Project DO1-400/18.12.2020 and the Academy of Finland grant nr. 320045 is gratefully acknowledged. 
This work was also been supported by Centros de Excelencia ``Severo Ochoa'' y Unidades ``Mar\'{\i}a de Maeztu'' program of the MCIN/AEI/ 10.13039/501100011033 (SEV-2016-0588, SEV-2017-0709, CEX2019-000920-S, CEX2019-000918-M, MDM-2015-0509-18-2) and by the CERCA institution of the Generalitat de Catalunya; by the Croatian Science Foundation (HrZZ) Project IP-2016-06-9782 and the University of Rijeka Project uniri-prirod-18-48; by the DFG Collaborative Research Centers SFB1491 and SFB876/C3; the Polish Ministry Of Education and Science grant No. 2021/WK/08; and by the Brazilian MCTIC, CNPq and FAPERJ.

% Author contributions
\newline\newline T.~Oka: MAGIC data analysis, paper drafting and edition;
T.~Saito: project leadership, MAGIC analysis cross-check, paper drafting and edition;
M.~Strzys: MAGIC analysis cross-check, paper drafting and edition.
The rest of the authors have contributed in one or several of the following ways: design, construction, maintenance and operation of the instrument(s) used to acquire the data; preparation and/or evaluation of the observation proposals; data acquisition, processing, calibration and/or reduction; production of analysis tools and/or related Monte Carlo simulations; discussion and approval of the contents of the draft.
\end{acknowledgements}
%--------------------------------------------------------------------

%--------------------------------------------------------------------
\bibliographystyle{aa}
\bibliography{main}
%--------------------------------------------------------------------

\begin{appendix}
\section{More detailed morphological investigations}
\label{section:residual}

\begin{figure}
  \centering
   \begin{tabular}{c}
    \begin{minipage}{1.0\hsize}
      \centering
      \includegraphics[width=\hsize]{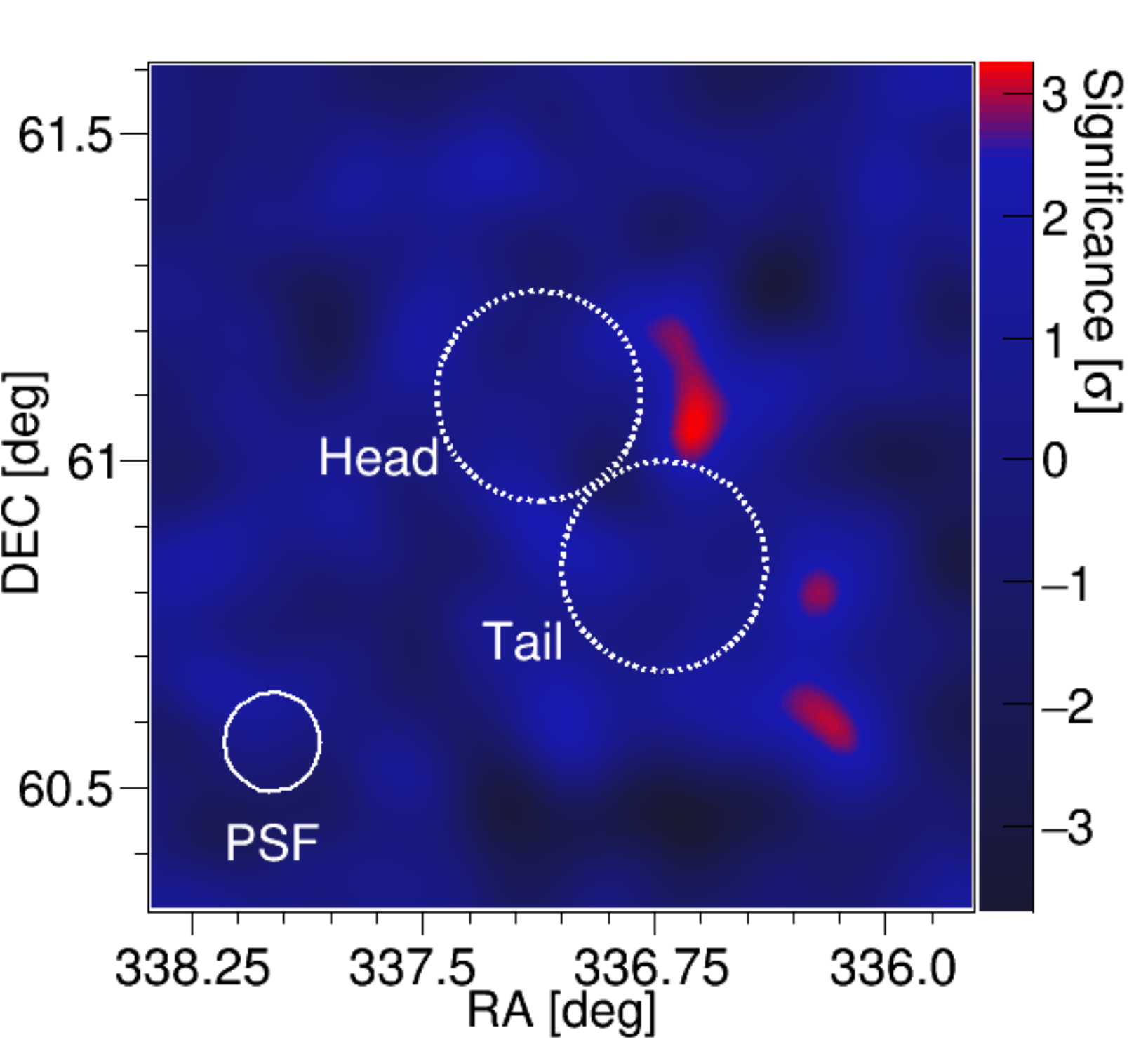}
    \end{minipage}
    \\
    \begin{minipage}{1.0\hsize}
      \centering
      \includegraphics[width=\hsize]{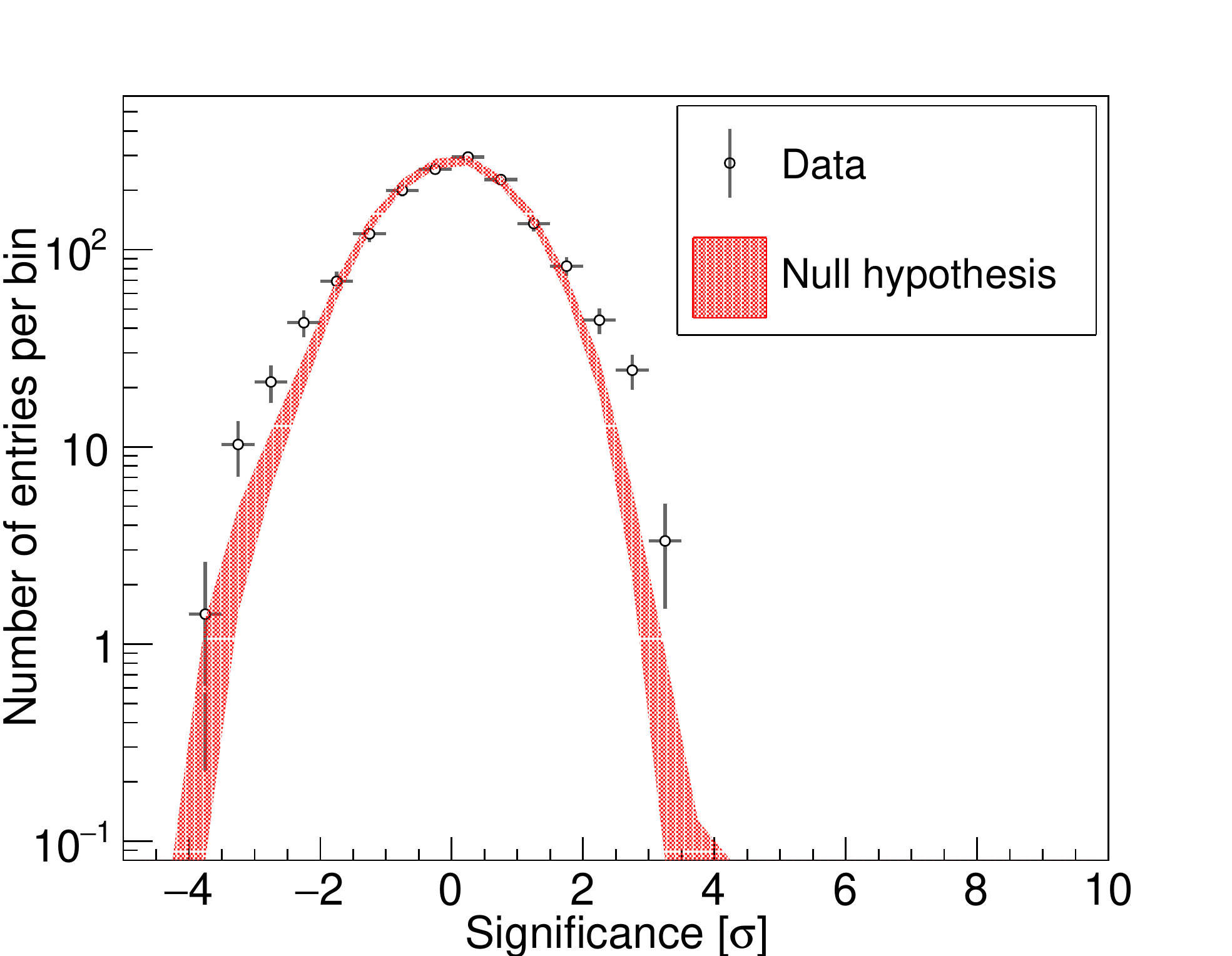}
    \end{minipage}
  \end{tabular}  
  \caption{
  Top: The residual map after subtracting two Gaussian sources in the energy range above 0.2~TeV.
  As for the Gaussian parameters, the locations are the same as the definition of {\it head} and {\it tail} and each $1\sigma$ extension radius is 0.085$^{\circ}$.
  Bottom: Pre-trial significance distribution of the residual map.
  }
  \label{fig:MAGIC_SkyMap_Residual}
\end{figure}

We used a double symmetric Gaussian function to examine the radiation peaks and select the analysis regions in the least biased way possible.
The best-fit parameters are as in Table~\ref{tab:SourceModel} for the center position, and 0.083$^{\circ}$ (0.087$^{\circ}$) for the $1\sigma$ extension of {\it head} ({\it tail}).
Fig.~\ref{fig:MAGIC_SkyMap_Residual} shows the residual map after subtracting two Gaussian sources and its significance distribution of the residuals.
The distribution is consistent with the null hypothesis, which indicates that, with the current statistics, the double Gaussian assumption is valid, though the true $\gamma$-ray source morphology may be more complex.

Although we cannot claim the proper source shape of the {\it head} and {\it tail} components from the present statistics, under the assumption that the source has Gaussian-like extension with $1\sigma$ of 0.085$^{\circ}$ (after removing the effect of PSF), the loss and contamination rate from the $\theta^{2}$ cut are estimated to be $23.5\%$ and $2.7\%$, respectively.
Further observations with better angular resolution could be helpful to determine a proper morphological model.

\section{Gas density in the emission regions}
\label{section:gasdensity}

\begin{figure*}
\centering
\includegraphics[width=\hsize]{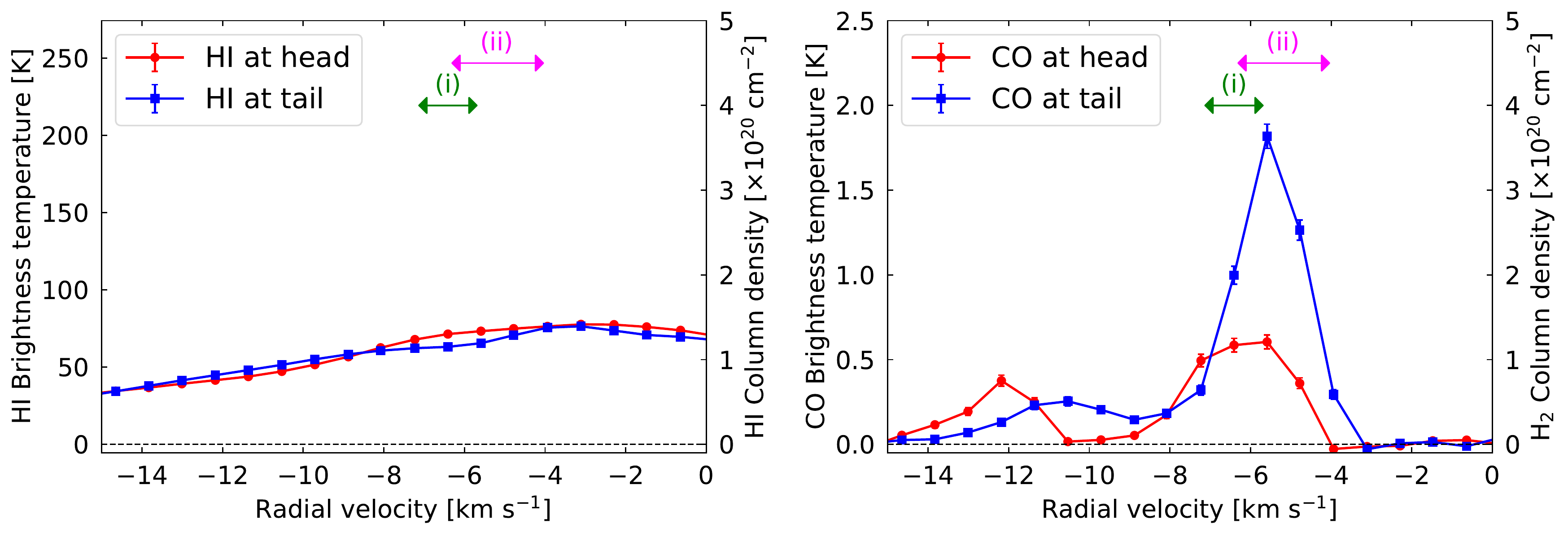}
\caption{The HI (left) and $^{12}$ CO ($J=1-0$) (right) radial profile at the {\it head} and {\it tail} region.
In both panels, red and blue data represent the profile of the {\it head} and {\it tail} regions.
The green arrow labeled (i) and magenta arrow labeled (ii) show the velocity ranges pointed out by~\cite{Kothes_2001} and \cite{Acciari:2009zz, Albert:2020ngw}, respectively.
}
\label{fig:RadialProfile}
\end{figure*}

We calculate the gas density in the two regions of SNR G106.3+2.7 with the following outline.
We use the data of HI line measured with the Dominion Radio Astronomy Observatory (DRAO) Synthesis Telescope~\citep{Landecker:2000jy} and $^{12}$CO ($J=1-0$) line measured with the Five College Radio Astronomy Observatory~\citep[FCRAO;][]{Heyer_1998} from the Canadian Galactic Plane Survey~\citep[CGPS;][]{Taylor_2003} database. These observations were carried out with the velocity resolution of 0.824~km~s$^{-1}$ at HI line and 0.98~km~s$^{-1}$ at CO line.
The following relationship is used to calculate the column density: $  N_{\mathrm{H}}\,[\mathrm{cm^{-2}}]= X \int^{v_{\mathrm{max}}}_{v_{\mathrm{min}}} T(v) dv$,
where $v$ is the radial velocity, $T(v)$ is the observed brightness temperature (K) and $X$ is the conversion factor~\citep{Dickey:1990mf}. HI-to-N$_{\rm HI}$ and CO-to-N$_{\mathrm{H_{2}}}$ are given by $X_{\mathrm{HI}} = 1.823\times10^{18}$~\citep{Dickey:1990mf} and $X_{\mathrm{CO}} = 2.0\times10^{20}$~\citep{Bolatto:2013ks}.
Fig.~\ref{fig:RadialProfile} shows the radial profiles of HI and $^{12}$CO ($J=1-0$) line.
There is a significant velocity dependence of the column density, especially in the CO data, which is a concern because the uncertainty of the velocity range affects the calculation of the gas density.
Here, we consider two cases on the velocity ranges that associates with SNR G106.3+2.7: (i) $-7.23$ to $-5.59$~km~s$^{-1}$ suggested by~\cite{Kothes_2001} and (ii) $-6.41$ to $-3.94$~km~s$^{-1}$ suggested by \cite{Acciari:2009zz, Albert:2020ngw}.
The clouds associated with the production of the observed $\gamma$-ray emission are assumed to be a spherical region around the emission center with a radius of $800\,\mathrm{pc} \times \mathrm{tan}(0.16^\circ) \sim 2.2\,\mathrm{pc}$ estimated from the MAGIC data as shown in~Table\ref{tab:SourceModel}.
The calculation results are summarized in Table~\ref{tab:Density}.
\begin{table}
  \centering
    \caption{Gas densities of hydrogen atoms at the {\it head} and {\it tail} region. $n_{\mathrm{HI}}$ and $n_{\mathrm{CO}}$ are estimated with the HI line and $^{12}$CO ($J=1-0$) line data, respectively.}
    \begin{tabular}{lcc}  \hline \hline
      Velocity range [$\rm{km~s^{-1}}$]          & $-7.23$ -- $-5.59$ & $-6.41$ -- $-3.94$  \\ \hline
      $n_{\mathrm{HI}}$ at {\it head} [cm$^{-3}$]  & 42 & 59 \\ 
      $n_{\mathrm{CO}}$ at {\it head} [cm$^{-3}$]  & 73 & 66 \\ 
      $n_{\mathrm{HI}}$ at {\it tail} [cm$^{-3}$]  & 38 & 55 \\ 
      $n_{\mathrm{CO}}$ at {\it tail} [cm$^{-3}$]  & 137 & 191 \\ \hline
    \end{tabular}
    \label{tab:Density}
\end{table}
There is not big difference of the results between the integration velocity ranges.
We use 100~cm$^{-3}$ and 200~cm$^{-3}$ as a gas density of {\it head} and {\it tail} regions for the modelling.

\end{appendix}

\end{document}